\documentclass[manuscript,screen]{acmart}

\AtBeginDocument{%
  }

\usepackage{tabularx}
\usepackage{multirow}
\usepackage{makecell}
\usepackage{graphicx}
\usepackage{booktabs}
\usepackage{rotating}
\usepackage{subcaption}
\usepackage{longtable}
\usepackage{hyperref}
\usepackage{tikz}
\usetikzlibrary{positioning, arrows.meta}
\usepackage{orcidlink}
\begin{document}

\title{Trust in Edge-Enabled IoT Security: Features, Challenges and Research Directions} 
\author{ESİN ECE AYDIN, \orcidlink{0000-0002-9299-0392}}
\affiliation{%
  \institution{Cyber Security and Privacy Research Lab., SPFLab,
Department of Computer Engineering,
Istanbul Technical University}
  \city{Istanbul}
  \country{Türkiye}
}
\email{aydinesi16@itu.edu.tr}

\author{ŞERİF BAHTİYAR, \orcidlink{0000-0003-0314-2621}}
\affiliation{%
  \institution{Cyber Security and Privacy Research Lab., SPFLab,
Department of Computer Engineering,
Istanbul Technical University}
  \city{Istanbul}
  \country{Türkiye}}
\email{bahtiyars@itu.edu.tr}

\author{GÜRKAN GÜR, \orcidlink{0000-0002-3105-4904}}
\affiliation{%
  \institution{Zurich University of Applied Sciences (ZHAW), Institute of Computer Science (InIT)}
  \city{Winterthur}
  \country{Switzerland}
}
\email{gurkan.gur@zhaw.ch}

\renewcommand{\shortauthors}{Aydin et al.}

\begin{abstract}
Providing autonomous intelligence, pervasive connectivity and usability to human life and industry has led to the emergence of the Internet of Things (IoT). To support time-sensitive and resource-constrained applications, IoT systems nowadays increasingly rely on edge computing. This brings computation and decision-making closer to end devices. In edge-enabled IoT architecture, latency and communication overhead are reduced, but interactions among a larger and more diverse set of devices, edge nodes, services, and data sources are introduced as well. In such environments, security and privacy mechanisms provide the foundation for protection, while trust management can assess the reliability of interacting entities and adapting secure decisions. In this paper, we systematically review the current state of trust management in edge-enabled IoT. To this end, we propose a comprehensive taxonomy that maps physical, network, and application architectural IoT layers against the consumer, commercial, industrial, and infrastructure IoT domains. We further investigate state-of-art research based on their trust design, how trust integrated into secure IoT operations, the attacks that effect trust management process. Based on these findings, we identify key gaps in current research and outline future directions for context-aware and adaptive trust management in edge-enabled IoT.
\end{abstract}



\maketitle

\section{Introduction}\label{sec:intro}

Security and privacy have long been central pillars of research in networked systems. However, securing IoT operations introduces unique challenges that security and privacy alone do not fully address, including reliable data exchange, dependable service provision, and context-aware decision-making \cite{chandrasekaran2024trust, hossain2024holistic}.
Modern IoT ecosystems no longer consist of just Internet-connected devices. Instead, IoT entities form dynamic relationships with neighboring devices, edge servers, cloud services, and autonomous applications. These heterogeneous entities continuously exchange information and rely on one another to complete distributed tasks and provide seamless services. Furthermore, next-generation IoT ecosystems are fundamentally characterized by the integration of edge AI. 
In this decentralized deployment, intelligence is distributed across edge and end devices rather than being centralized in the cloud. By this change, privacy and security enforcement are also distributed across heterogeneous edge and end nodes \cite{Bounaira2025Trustworthy,Feng2023Blockchain}. However, they do not by themselves determine with whom an entity should interact, whose data or services should be relied upon, or how much autonomy should be granted in day-to-day IoT operation. 
Therefore, trust emerges as an equally critical third dimension. According to \cite{alhandi2023trust}, trust enables IoT systems to assess the reliability of relationships between things and realize the full potential of their interactions.  From consumer-related ones to infrastructure-wide IoT applications, integrating trust mechanisms is an important aspect for the successful operation of IoT. Besides, as stated in \cite{Verma2023RepuTE}, trust assessment can make a substantial impact on the overall security of IoT systems.
In edge-enabled IoT, trust management are not centralized: models, training data, and score updates can be evaluated by end nodes themselves. This enables privacy-preserving decisions, reduced latency and bandwidth usage. However, local decision making and cross-layer collaboration become important design issues in trust management of next-gen IoT as mentioned by \cite{Ranathunga2021The}.

Although trust is recognized as a one of the fundamental requirements for reliable IoT operation \cite{najib2019survey, chandrasekaran2024trust}, currently, trust remains is an abstract concept without a widely accepted definition. 
The literature offers a wide spectrum of definitions: ranging from subjective confidence and probabilistic behavioral belief \cite{Mohanta2025Smart, Li2024Blockchain, Wang2020MTES, Aaqib2024Behavior}, to reputation-based and QoS-driven trustworthiness \cite{ Guo2022Endogenous, Verma2023RepuTE}, and to multi-dimensional, context-aware, or AI-predicted trust \cite{chandrasekaran2024trust, Bampatsikos2025Trust, Messina2025A, Ranathunga2021The, sagar2024understanding}.
The literature lacks a consistent view of whether trust concerns devices, data, services, or all three, resulting in fragmented methodologies and inconsistent evaluation.

This survey addresses this gap by examining trust in conjunction with security and privacy. 
It examines the definition of trust in the literature, the processes of trust management, and its integration into IoT operations to improve resilience, robustness, and secure decision-making. 
This survey is structured around the following research questions: 
\begin{itemize}
  
    \item \textbf{RQ1 -- } How is trust management in edge-enabled IoT characterized across application domains, architectural layers, and trust targets (device, data, and service), and what recurring design patterns emerge in trust management?

    \item \textbf{RQ2 -- } How do reviewed schemes use trust scores to drive IoT operational decisions, and trust-related attacks, including routing, reputation, data-integrity, and identity-related threats are targeted?

    \item \textbf{RQ3-- } What research gaps remain in IoT trust management, and what future research directions are needed for scalable and resource-efficient trust mechanisms, adversary-aware and attack-adaptive models in emerging IoT environments including edge-enabled, agent-based, immersive, and physically–virtually coupled deployments?

\end{itemize}

\subsection{Existing Surveys} \label{sec-intro-surveys}
We compare eighteen recent survey papers published between 2019 and 2026 with our work, as given in Table \ref{tab:survey-comparison}. The comparison covers several aspects of trust management. Trust target indicates trust subjects each survey focuses. Trust design taxonomy indicates whether the survey provides a structured trust-management design, including evidence acquisition, trust modeling, dissemination, and maintenance. Operational roles of trust capture how trust is used in practice to support IoT decisions, such as access control, secure routing, intrusion detection, service selection, or data aggregation. Trust attack analysis indicates whether trust-related attacks are systematically identified and analyzed. The last one, architecture scope, gives the IoT deployment targeted by each survey.

\begin{table*}[ht]
\centering
\footnotesize
\caption{Comparison of Existing Surveys on Trust Management in IoT with Our Work}
\label{tab:survey-comparison}
\resizebox{\textwidth}{!}{%
\begin{tabular}{l c c c c c c c c}
\toprule
\textbf{Survey} & \textbf{Year}
& \multicolumn{3}{c}{\textbf{Trust Targets}}
& \textbf{Trust Design}
& \textbf{Operational}
& \textbf{Trust Attack}
& \textbf{Architecture}
\\
\cmidrule(lr){3-5}
& & Device & Data & Service & \textbf{Taxonomy} & \textbf{Roles of Trust} & \textbf{Analysis}
& \textbf{Scope}
\\
\midrule

Xu et al.~\cite{xu2026Enhancing}                            & 2026 & \checkmark & \checkmark & $\sim$     & \checkmark    & $\sim$ & \checkmark & Edge-Enabled IoV 
\\ \hline

Wang et al.~\cite{wang2026trust}                             & 2026 & \checkmark &   $\sim$   & $\sim$ & \checkmark   &     & \checkmark    & Distributed IoV      
\\\hline

D'Aniello \& Fotia~\cite{d2025blockchain}                    & 2025 & \checkmark &      & $\sim$     & $\sim$      &      &     & Edge-Enabled IoT 
\\\hline

Jmal et al.~\cite{jmal2025blockchain}                         & 2025 & \checkmark &    & \checkmark     &  $\sim$   &      & \checkmark     & SIoT      
\\\hline

Kamble et al.~\cite{kamble2024research}                       & 2024 & \checkmark &            &            & \checkmark    &            & $\sim$     & General IoT 
\\\hline

Sagar et al.~\cite{sagar2024understanding}                    & 2024 & \checkmark &            & \checkmark & \checkmark    & \checkmark & \checkmark     & SIoT                  
\\\hline

AlMarshoud et al.~\cite{almarshoud2024security}                 & 2024 & \checkmark &     & $\sim$     & \checkmark        &     & $\sim$     & Decentralized IoV   
\\\hline

Tyagi et al.~\cite{tyagi2023detailed}                         & 2023 & \checkmark &            &            & \checkmark    & \checkmark & \checkmark     & WSN-Based IoT  
\\\hline

Shirvani \& Masdari~\cite{shirvani2023survey}                 & 2023 & \checkmark &   \checkmark           &            & $\sim$        & \checkmark & $\sim$     & General IoT    
\\\hline

Aaqib et al.~\cite{aaqib2023iot}                              & 2023 & \checkmark &  $\sim$     &            & \checkmark    &            &      & General IoT 
\\\hline

Lenard et al.~\cite{lenard2023exploring}                      & 2023 & \checkmark &            & $\sim$     & \checkmark    & $\sim$     & \checkmark & Distributed IoT   
\\\hline

Alhandi et al.~\cite{alhandi2023trust}                        & 2023 & \checkmark &  \checkmark    &            & \checkmark    &            & $\sim$     & WSN-Based IoT      
\\\hline

Wang et al.~\cite{wang2022survey}                              & 2022 & \checkmark & $\sim$     & $\sim$     & \checkmark    & \checkmark & \checkmark    & Heterogeneous IoT  
\\\hline

Konsta et al.~\cite{konsta2022trust}                          & 2022 & \checkmark &     $\sim$        &            & \checkmark        &            & $\sim$     & Edge-Enabled IoT 
\\\hline

Muzammal et al.~\cite{muzammal2020comprehensive}              & 2021 & \checkmark & \checkmark & $\sim$  & \checkmark    &      & \checkmark     & General IoT   
\\\hline

Marche \& Nitti~\cite{marche2020trust}                        & 2020 & \checkmark &            & \checkmark & $\sim$        & $\sim$     & \checkmark   & SIoT                  
\\\hline

Pourghebleh et al.~\cite{pourghebleh2019comprehensive}      & 2019 & \checkmark &  \checkmark   &     & $\sim$        & $\sim$     & $\sim$     & General IoT 
\\\hline

Najib et al.~\cite{najib2019survey}                           & 2019 & \checkmark &            &            & \checkmark    &            &            & General IoT 
\\

\midrule
\textbf{Our work}                                             & 2026 & \checkmark & \checkmark & \checkmark & \textbf{\checkmark} & \checkmark & \checkmark
& Edge-Enabled IoT     
\\

\bottomrule
\end{tabular}%
}

\footnotesize{\checkmark{} = explicitly covered, $\sim$ = partially covered, blank = not covered}
\end{table*}

Across all the surveys examined, device trust is the only target treated consistently. This reflects that IoT trust management has device-centric model, where IoT nodes, sensors are evaluated for their trustworthiness. While a few recent surveys have begun to expand their scope such as \cite{shirvani2023survey, alhandi2023trust, muzammal2020comprehensive, pourghebleh2019comprehensive} including data trust, and \cite{jmal2025blockchain, sagar2024understanding,marche2020trust} focusing on service trust which is unsurprising given their focus on Social IoT and service recommendation or selection. 
To the best of our knowledge, no prior survey paper covers all three trust targets; the closest cases are the review by \cite{xu2026Enhancing,wang2026trust,wang2022survey,muzammal2020comprehensive}, however, they still address one or two of the remaining trust targets only partially.

According to \cite{sagar2024understanding}, trust management schemes are categorized into recommendation-based, reputation-based, prediction-based, and policy-based groups. However, this classification combines two different aspects into a single taxonomy: recommendation and reputation-based schemes describe the source of trust evidence, whereas prediction and policy-based schemes describe how trust is computed. Since these dimensions are independent, a single scheme may belong to multiple categories simultaneously. Therefore, in this survey, trust evidence sources and trust computation mechanisms are treated as separate dimensions.
\cite{aaqib2023iot, najib2019survey} classify trust models across a broad architectural scope, ranging from centralized to distributed IoT systems. However, these surveys do not analyze the operational roles of trust or trust-related attacks. \cite{pourghebleh2019comprehensive, muzammal2020comprehensive} provide more detailed discussions of trust computation methods and adversarial threats. \cite{kamble2024research} frames trust management as a core enabler of secure IoT applications across five main domains: agriculture, education, healthcare, home security, and transportation. The authors further organize prior work using a five-part taxonomy, including trust measures (QoS, social-based trust), trust circulation (centralized, distributed), trust algorithms (fuzzy logic, machine learning, Bayesian methods, entropy-based models, regression), trust generation (single, multi attribute), and highlighting trust-related attacks. However, the study lacks an evaluation of the operational roles of trust within IoT systems. While \cite{shirvani2023survey} categorizes the operational roles of trust, it does not offer any taxonomy for trust management design, and it omits giving any structure how trust is addressed in state-of-the-art approaches. Edge-enabled IoT architectures are examined in \cite{xu2026Enhancing,d2025blockchain,konsta2022trust} where trust is maintained distributed across devices, fog/edge nodes, and cloud rather than confined to a centralized structures. They emphasize decentralized mechanisms such as blockchain, and edge-local reputation. \cite{konsta2022trust} reviews state-of-the-art approaches together with the threat models. The authors categorize 53 papers and evaluate each one on seven dimensions: information gathering, trust formation, propagation and update, threat model, experiments, and simulator. 
Different taxonomies are suggested based on layers, deployment modes, algorithms, or enabling technologies. However, the taxonomies in \cite{d2025blockchain, jmal2025blockchain, shirvani2023survey, marche2020trust, pourghebleh2019comprehensive} remain incomplete with respect to the full trust lifecycle.

\subsection{Contributions \& Paper Organization} \label{sec-intro-contributions}

Building on the the research questions outlined in Section \ref{sec:intro} and comparison in Table \ref{tab:survey-comparison}, this survey makes the following contributions:
\begin{itemize}
    \item We present a systematic literature review of trust-management techniques in edge-enabled IoT and classify the selected studies. We use a taxonomy that maps each study across IoT layers and four application domains: consumer, commercial, industrial, and infrastructure IoT.

    \item We provide an analysis of trust-related attacks by relating adversarial objectives, attacker behavior patterns, and affected stages of the trust-management process. We further examine how trust-management outputs are used to inform decisions and trigger actions within IoT systems.

    \item We identify key research challenges in building trustworthy edge-enabled next-generation IoT. Then, we outline possible future research directions.
\end{itemize}

The remainder of this paper is organized as follows. First, Section~\ref{sec-foundations} provides background, gives definitions, and explains trust management process along the trust components. Next, Section~\ref{sec-methodology} presents the systematic review methodology we follow for comparative analysis. 
Then, representative state-of-the-art frameworks are analyzed and discussed in Section~\ref{sec-current-frameworks}. 
Section~\ref{sec-road-ahead} discusses open research challenges, including architectural, trust-model, adversarial, and ecosystem barriers, and outlines future directions. Finally, Section~\ref{sec-conc} concludes the paper.

\section{Trust, Security and Privacy Foundations in IoT} \label{sec-foundations}

Security in IoT concerns protecting devices, communication channels, and services against unauthorized access, manipulation, and disruption with confidentiality, integrity, and availability as its core objectives. However, IoT deployments are heterogeneous, resource-constrained, and large-scale which makes conventional security mechanisms difficult to apply \cite{Deng2024Lightweight}. Static and heavyweight cryptographic protocols suffer from significant latency \cite{Din2025Building, Fang2020Fast}, energy consumption \cite{Chaganti2025A}, and may be vulnerable in key management \cite{Fang2020Fast}. 
As a result, IoT security research seeks lightweight protection and detection mechanisms.
The central challenge in this landscape is the transition from perimeter security to zero-trust models, where every entity must continuously prove its integrity and trustworthiness.

Privacy encompasses protecting user identity, preventing personal inferences, and ensuring regulatory compliance. IoT devices continuously collect data about individuals, environments, and activities, often without the user's explicit knowledge. Consequently, privacy risks arise. Protection mechanisms such as data anonymization, differential privacy, encryption, and access control can be used to limit exposure. However, deploying these in IoT requires careful consideration. For example, when IoT systems are deployed by static role-based policies, access rights cannot adapt as IoT environment changes. Besides, cryptography also poses challenges in key lifecycle management. Informed consent, data ownership, and accountability become harder to sustain across multi-party, cross-layer data flows.
Accordingly, recent research integrates trust-based differential-privacy-based mechanisms \cite{Nawshin2024AI}, federated learning (FL) for sensitive data \cite{Khan2025An}, and privacy-preserving access control \cite{Gheisari2025A}.
The central challenge in this landscape is the transition from data secrecy to context-aware privacy preservation in order to balance data utility with user privacy. 

\subsection{IoT System Characteristics} \label{sec-foundations-architecture}

Securing IoT deployments is inherently difficult: devices are heterogeneous and resource-constrained, adversaries may be external or already authenticated, and privacy must be preserved while data is continuously sensed, transmitted, and processed. The literature characterizes IoT systems along several complementary dimensions. IoT systems are commonly described through functional layers. These are typically physical (perception), network, and application layers \cite{sun2025survey, tyagi2023detailed, alhandi2023trust, konsta2022trust}, with extended variants with processing or transport layers \cite{chandrasekaran2024trust}. IoT systems can also be characterized by the location of sensing, computation, and decision-making along the device–edge–fog–cloud continuum. These are device-level \cite{Turkina2020Approach, Lakhan2024Augmented}, edge-level, fog-level \cite{Rahman2023Using}, cloud-level \cite{Kumari2020Blockchain} and device–edge–cloud combinations \cite{Souri2024A, Sun2024Trust, d2025blockchain, Messina2025A}. Third, IoT systems can be further distinguished by their application domains. In order to reflect requirements of the deployed systems, these could be consumer, enterprise, commercial, industrial, or military application domains. For example, \cite{nashiruddin2019coverage} investigates IoT application domains under two: critical IoT and massive IoT. 
Additional dimensions could be the organizational architecture of IoT entities, such as centralized, hierarchical, or peer-to-peer structures, as well as node mobility, ranging from static infrastructure to mobile vehicular and aerial networks \cite{Dong2023Securing, Zhang2024A, Chen2024UITDE}.

\begin{figure}[h]
    \centering
    \includegraphics[width=\linewidth]{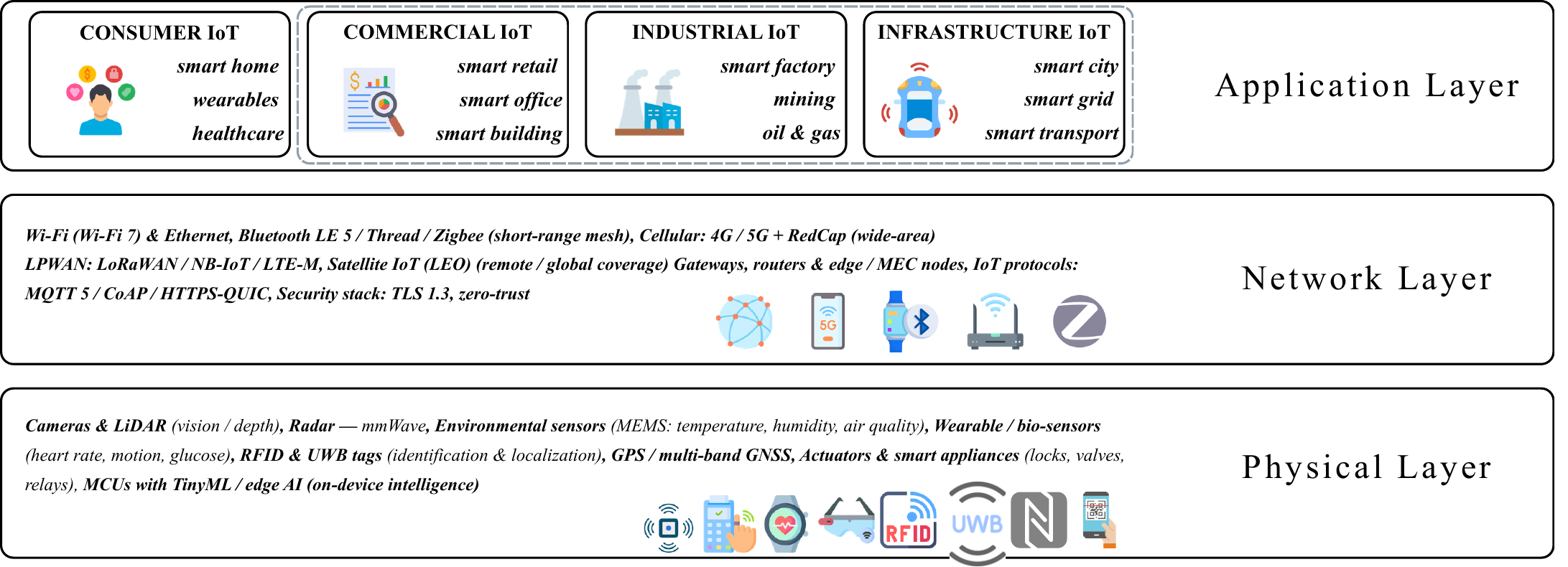}
    \Description{Three-Layer Model of IoT Systems Across Consumer, Commercial, Industrial, and Infrastructure Domains}
    \caption{Three-Layer Model of IoT Systems Across Consumer, Commercial, Industrial, and Infrastructure Domains}
    \label{fig:iot-layers}
\end{figure}

For this survey paper, we consider edge-enabled IoT architectures based on a three-layer model and four types of application domains, as presented in Figure~\ref{fig:iot-layers}. We divide IoT application domains by end-user and market as Business-to-Consumer (B2C) and Business-to-Business (B2B). The B2C segment includes consumer IoT which encompasses devices and ecosystems designed for direct use by individuals to improve quality of life, convenience, and personal wellness. The B2B segment is significantly more diverse, focusing on enterprise-level applications where IoT is deployed to optimize processes, reduce costs, and manage assets. It includes commercial, industrial, and infrastructure IoT. 

\subsection{Trust in IoT} \label{sec-foundations-definition}

The trust landscape defines who can securely and reliably operate on behalf of whom, encompassing trust in devices, the data they produce, and the services that act upon that information. It determines not only who is allowed to participate in the system, but also how much autonomy each participant is granted under varying conditions. 
Many researchers associate trust management solely with social IoT environments \cite{Moeinaddini2025A, Narang2021A, Latif2022A, Magdich2022A}. In reality, devices in all IoT systems depend on many others for data and services. A well-known example of this occurred in February 2020 when Simon Weckert artificially created a virtual traffic jam on Google Maps by transporting 99 smartphones in a wheelbarrow \cite{marche2020trust}. Although not a conventional IoT deployment, this incident shows how untrusted data can propagate through a system and lead to incorrect operational decisions.
To better understand trust in IoT operations, it is first necessary to establish key trust-related definitions and then examine how trust management is performed in IoT systems. Trust management schemes identify three participant roles: the trustor, who evaluates and holds trust; the trustee, whose trustworthiness is assessed; and the recommender, who supplies indirect evidence about a third party \cite{Khoshvaght2025An}. These roles provide the basis for defining the key concepts underlying trust assessment in IoT. 
We additionally use the following definitions:

\textit{Definition 1:} Trustworthiness refers to the degree to which a trustee exhibits observable qualities such as reliability, integrity, competence, and benign behavior. \cite{Tsekenis2025Flexible} further incorporates factors such as capacity, energy, performance, and cost. In general, trustworthiness denotes the inherent properties of the trustee that justify a given trust assessment.

\textit{Definition 2:} Trust is the quantitative or qualitative measure of the confidence that a trustor $a$ places in a trustee $x$ to behave as expected  over interval $[t_0,t]$. Formally, we can define trust as in Equation \ref{eq.trust} where $E_a(x)$ denotes the trust evidence available to $a$, and $f(\cdot)$ is the trust evaluation function. 
\begin{equation} \label{eq.trust}
T_a(x \mid t) = f\!\left(E_a(x)\right),
\end{equation}

\textit{Definition 3:} Device trust is the value $T_a(x \mid t) \in [0,1]$ by which trustor $a$ expresses confidence that trustee device $x$ will operate benignly up to time $t$, computed from evidence $E_a$ about $d$’s behavioral history, operational integrity, and adherence to expected node-level functions.

\textit{Definition 4:} Data trust is the graded confidence $T_a(x \mid  t)$ by which trustor $a$ judges that trustee data $x$ is reliable for use, based on evidence $E_a(x)$ about its accuracy, integrity, or freshness.

\textit{Definition 5:} Service trust is the graded confidence $T_a(x \mid t)$ by which trustor $a$ judges that service $x \in \mathcal{S}_x$ (all services of a specific IoT device) will deliver its advertised functionality correctly, completely, and within agreed quality-of-service over interval $[t_0, t]$, without malicious or negligent behavior, based on service-specific evidence $E_a(x)$.  

\textit{Definition 6:}  A trust relationship is a measurable, subjective, and transferable association \cite{Yang2023Blockchain} between a trustor and a trustee, representing the trustor’s context-dependent confidence in the trustee.

\textit{Definition 7:} Trust-related attacks 
are a class of adversarial strategies in which malicious entities deliberately manipulate the inputs, computation, or dissemination of trust scores to subvert the trust management system itself. Unlike conventional network attacks that target data or availability, trust poisoning attacks manipulate or corrupt the evidence upon which trust assessments are built. 

\begin{figure}[h]
    \centering
    \includegraphics[width=\linewidth]{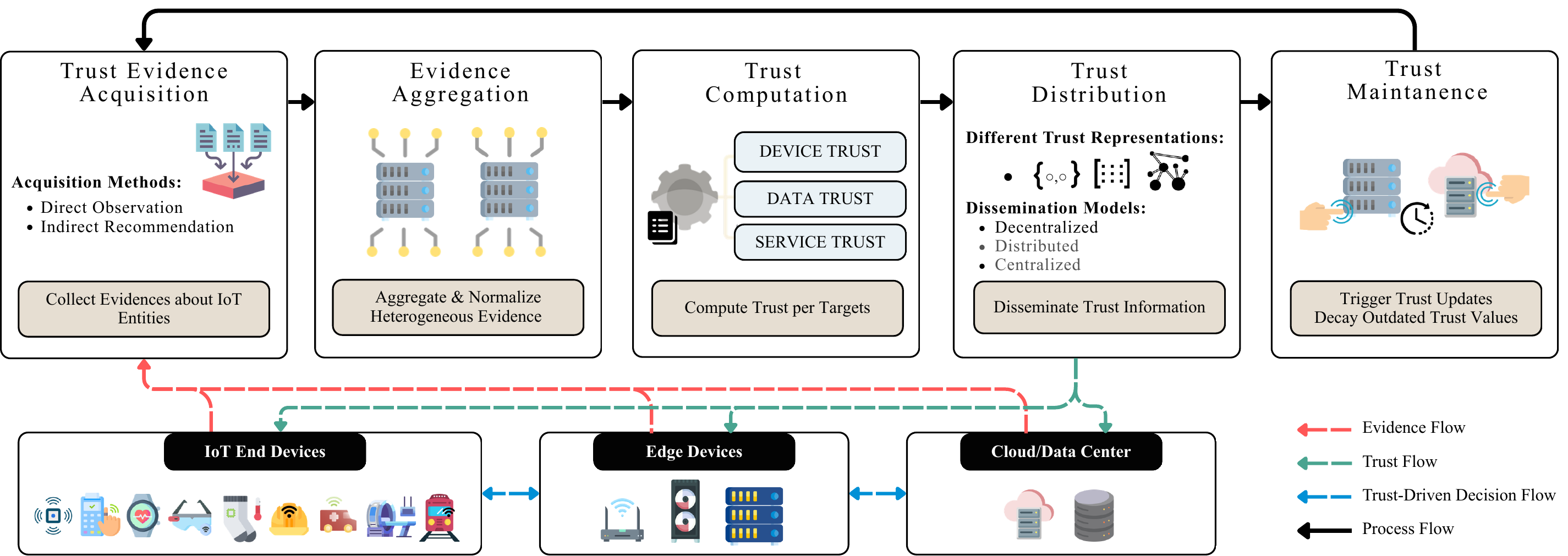}
    \Description{Conceptual Trust Management Process}
    \caption{Conceptual Trust Management Process}
    \label{fig:trust-management-process}
\end{figure}

In order to develop trust relationships and protect against adversarial behaviors and cyber-physical attacks, trust must be managed continuously. Figure \ref{fig:trust-management-process} illustrates the conceptual workflow of trust management process in IoT.
Trust management begins with \emph{evidence acquisition}, in which a trustor may collect device, data, and service related information that serve as inputs to trust assessment. These evidences are then processed during \emph{evidence aggregation}, where an initial trust value is formed, direct and indirect evidence may be aggregated, and normalized. After that, a trust modeling technique (rule-based, probabilistic, ML, or blockchain-assisted) computes a trust value per targets in \textit{trust computation} step. 
Subsequently, \textit{trust distribution} determine where trust is calculated, stored, and shared across the network. For example, at an edge device these trust values can be stored and how updated values are propagated to other entities. Trust scores can be then used by IoT end devices, edge nodes or cloud/data center to support decisions in IoT operations such as routing, access control, service selection, and intrusion detection. Finally, in \textit{trust maintenance} trust scores are updated after new interactions or elapsed time.

Most IoT trust management frameworks follow a common process, as illustrated in Figure~\ref{fig:trust-management-process}. Before discussing current frameworks, it is important to briefly discuss some important components of this process.

\subsubsection{Trust Evidence Acquisition} \label{sec-state-acquisition}
It is a process that an IoT system uses to gather information about IoT entities.
Some surveys distinguish single and multi attribute evidence collection \cite{sagar2024understanding,kamble2024research,alhandi2023trust,konsta2022trust}. We omit this distinction because most state-of-the-art frameworks use multiple attributes; instead we classify evidence acquisition by source: direct or indirect. Direct evidence refers to trust information based on an assessing entity ~$a$'s first-hand observations of a target~$x$, which may be a device, data item, stream, or service. Let $\tau(x) \in \{\mathrm{device}, \mathrm{data}, \mathrm{service}\}$ denote the trust target type of~$x$. Direct evidence is represented by a pair of positive and negative outcomes, $(p_{a,x}, n_{a,x})$, and a vector of normalized metrics ${m}_{a,x} = (m_{a,x}^{(1)}, \ldots, m_{a,x}^{(L)})$, where each $m_{a,x}^{(\ell)} \in [0,1]$. The direct trust score is then computed uniformly as in Equation \ref{eq:direct-trust-unified} where $\Phi(\cdot)$ is an aggregation function whose inputs are interpreted according to $\tau(x)$. 
\begin{equation}
    T_d(a,x) = \Phi\!\left(p_{a,x},\; n_{a,x},\; {m}_{a,x};\; \tau(x)\right)
    \label{eq:direct-trust-unified}
\end{equation}
In practice, this reduces to either a success ratio as shown in Equation \ref{eq:direct-trust-ratio}:
\begin{equation}
    T_d(a,x) = \frac{p_{a,x}}{p_{a,x} + n_{a,x}}
    \label{eq:direct-trust-ratio}
\end{equation}
or a weighted metric combination as given in Equation \ref{eq:direct-trust-weighted}:
\begin{equation}
    T_d(a,x) = \sum_{\ell=1}^{L}  \omega_{\ell}\!\left(\tau(x)\right) \cdot m_{a,x}^{(\ell)}, \quad \sum_{\ell=1}^{L} \omega_{\ell}\!\left(\tau(x)\right) = 1
    \label{eq:direct-trust-weighted}
\end{equation}

Direct evidence can be further divided into two forms, as shown in Figure \ref{fig:trust-design-dimensions}: experience and observations. Nodes' direct interactions with each other provide experience-based evidence, such as recording the number of successfully forwarded packets relative to failed ones \cite{Osamy2025TAGSCS}. Observation-based evidence, on the other hand, is obtained through monitoring or overhearing interactions between other nodes.
Direct evidence acquisition is generally considered more objective observation from the perspective of an IoT node. It is less dependent on other's subjective observations, therefore it is less vulnerable to trust poisoning attacks.
Although indirect evidence is more susceptible to trust-poisoning attacks, it becomes essential when direct interactions are unavailable. In such cases, trust is inferred from recommendations or reputation (consensus). Recommendations represent evaluations shared by neighboring nodes regarding the reliability of a target entity \cite{marche2020trust}.
In contrast, reputation (consensus) aggregates trust opinions from multiple nodes into a community-level score. However, its effectiveness depends on the credibility of obtained trust value \cite{Yang2023Blockchain} and recommenders \cite{Ahmed2022Link}.
The set of all indirect source can be denoted as $\mathcal{R}_{x}$, and each source $r$ providing opinions about target~$x$. $m_{r}(x) \in [0,1]$ denotes the normalized trust opinion that $r$ contributes about~$x$. Indirect trust can be computed as in Equation \ref{eq:indirect-trust} where $\omega_{r} \in [0,1]$ is the credibility weight assigned to source~$r$:
\begin{equation}
    T_r(x) = \frac{\sum_{r \in \mathcal{R}_{x}} \omega_{r}\, m_{r}(x)}{\sum_{r \in \mathcal{R}_{x}} \omega_{r}}
    \label{eq:indirect-trust}
\end{equation}

In hybrid trust evidence acquisition models, direct and indirect evidence are combined to obtain an aggregated trust score. The trust that assessing entity~$a$ assigns to target~$x$ is commonly computed as a weighted combination \cite{Wang2025A} where $\alpha \in [0,1]$ controls the relative influence of direct versus indirect evidence:
\begin{equation}
    T(a,x) = \alpha \cdot T_d(a,x) + (1 - \alpha) \cdot T_r(x)
    \label{eq:trust-hybrid}
\end{equation}

This hybrid formulation is motivated by a common limitation of IoT deployments: direct evidence is often incomplete, intermittent, or outdated by the time a stable trust estimate can be formed. IoT nodes try to collect signals related to competence, honesty, and security \cite{Bangui2025Leveraging}. However, building a reliable interaction history remains a significant challenge \cite{Messina2025Forming}. When first-hand interaction history is limited, schemes typically reduce weight $\alpha$ and rely more heavily on recommendations or reputation \cite{Sharma2024Multi, Guo2022ITCN}; as sufficient direct observations accumulate, $T_d(a,x)$ is given greater weight and dependence on potentially unreliable indirect sources is reduced. 
On the other hand, another challenge is that IoT environments often change faster than the evidence base used to support trust decisions \cite{Khan2025An,  Yu2023CET}. Sparse interactions therefore yield insufficient evidence for stable trust relationships. To compensate, some studies incorporate interaction intensity as an additional trust parameter \cite{Sharma2025Evaluation}, while \cite{marche2020trust} introduces a relationship component that assigns the highest value to directly connected entities and, for indirect pairs, derives trust from the weakest link along the available social path.
However, if sufficient direct observations and trustworthy recommendations are both unavailable, many existing schemes simply assign a default initial trust score to previously unknown entities, which may lead to inaccurate trust decisions during early interactions \cite{Huang2025Towards}. 

\subsubsection{Trust Computation} \label{sec-state-computation}

Trust computation maps aggregated evidence to a trust value for a given target, such as a device, data, or service. In the state-of-the-art, there are many families of computation methods, each differing in how evidence is interpreted, combined, and converted into a trust value.

Rule-based and weighted-sum models combine selected trust inputs using predefined rules, weights, thresholds, or ranking methods. They are generally lightweight and interpretable, but their effectiveness depends on how the rules and weights are defined~\cite{Bampatsikos2025Trust, Ali2024TIHCS}. Fuzzy-logic models use expert-defined rules to handle imprecise inputs and avoid relying on strict thresholds~\cite{Rehman2025A}. Probabilistic and statistical approaches model trust as uncertain or evolving evidence and update it as new observations become available. Examples include Beta-based reputation, Bayesian updating, and Markov models~\cite{Alsheakh2020Towards}. Subjective logic further represents belief, disbelief, and uncertainty separately, rather than reducing them to a single value. Learning-based approaches, including neural networks (NN), clustering, and federated learning, learn trust patterns from behavioral or interaction data and can capture relationships that are difficult to express through predefined rules~\cite{Moeinaddini2025A, Zhang2021AIT, Hammad2025Anomaly}. Agent-based designs also appear as multi-agent reputation and clustering schemes. They treat IoT nodes as cooperating agents whose group membership follows local trust~\cite{Fortino2023A}, while multi-agent reinforcement learning uses interacting learners to update trust-aware decisions at runtime~\cite{Tuncel2025SAFE, Chen2023A}. However, learning based methods generally introduce higher data and computational requirements. Game-theoretic approaches model trust-related interactions as strategic behavior between entities and can incorporate mechanisms such as voting, incentives, and coalition formation~\cite{Moudoud2022Detection}. Blockchain is also used in trust computation frameworks, mainly to provide a distributed mechanism for recording, sharing, and protecting trust values rather than serving as a trust computation method by itself~\cite{Latif2023MarketTrust, Sinha2024AI}. No single approach is universally preferred: the choice depends on the trust target (device, data, or service), the available evidence, and the resource constraints of the deployment \cite{ akli2023survey}. Besides, authors in \cite{Bampatsikos2025Trust} highlight the need for trust modeling while considering which IoT domains it will be adapted to.

\subsubsection{Trust Distribution} \label{sec-state-distribution}
Trust distribution is the stage at which computed scores are stored and shared so that other entities can use them for operational decisions. Before scores can be disseminated, they must be represented in a comparable form. In the state-of-the-art, trust may be modeled as a binary, nominal, discrete, or continuous value~\cite{akli2023survey}. Most schemes use a scalar score in $[0,1]$, while others keep multiple trust dimensions or map entities to discrete states such as trustworthy or malicious~\cite{Ismail2024Towards, Kuppusamy2024IoT, Khan2025An, Kumar2025AI, Wang2024Secure, Ali2024TIHCS}. Trust can also be represented as a vector, a matrix, or a graph of relationships, which enables indirect inference when only partial evidence is available~\cite{Messina2025A, Rajendran2022Friendliness}. 
Once a representation is chosen, two further questions arise: where trust scores are stored, and how they are disseminated across IoT network. Storage may be local to the evaluating node, held at an edge or fog gateway, or recorded on a shared ledger. Dissemination then determines who receives those stored values and along which path: top-down, horizontally among peers, through cluster heads, or by consensus exchange. IoT trust management can be implemented through centralized, distributed, or decentralized architectures. Among the surveyed studies, decentralized architectures are our focus. Table~\ref{tab:arch-comparison} compares the three families in terms of computation placement, dissemination strategy, typical use cases, and key drawbacks.
\begin{table}[htbp]
\centering
\caption{Architectural Comparison of Trust Management Placement}
\label{tab:arch-comparison}
\footnotesize
\begin{tabular}{@{} p{2cm} p{3cm} p{3cm} p{3cm} p{2.5cm} @{}}
\toprule
\textbf{Architecture} & \textbf{Computation} & \textbf{Dissemination} & \textbf{Best Suited For} & \textbf{Key Drawback} \\
\hline
\textbf{Decentralized (Our Focus)} &
Edge compute locally; cloud syncs globally &
Local gateway push; regional cloud coordination &
Large-scale, mixed-resource IoT (e.g., IIoT, smart cities) &
Complex gateway management \\  \hline
Distributed &
Peer evaluation without central coordination&
P2P broadcast or consensus protocols &
Dynamic, ad-hoc networks (e.g., VANETs) &
High resource overhead on constrained sensors \\ \hline
Centralized &
Central server processes all raw telemetry, compute trust &
Cloud broadcasts lists/policies to nodes &
Static, low-power networks with stable links &
Single point of failure; high latency \\ 
\bottomrule
\end{tabular}
\end{table}

In centralized architectures, central controller collects interaction evidence, computes trust, and disseminates the result top-down through blacklists, trust tokens, or policy updates. This model simplifies management and benefits from global visibility, but it introduces latency, a single point of failure, and scalability bottlenecks when all raw telemetry must be uploaded for analysis~\cite{Xiong2023BDIM}. In decentralized architectures, computation is shifted to intermediate nodes such as edge gateways, fog brokers, or cluster heads. These nodes evaluate trust locally, aggregate neighborhood reputation, and share summaries horizontally with peer managers while synchronizing with upper tiers.. In distributed architectures, there is no fixed trust coordinator: individual nodes evaluate peers directly and propagate updates through broadcast or consensus protocols. This supports ad-hoc and vehicular environments, but it imposes additional communication and computation costs on constrained devices.

\subsubsection{Trust Maintenance} \label{sec-state-update}

Trust assessments made at the beginning should not be considered permanent. Trust must be maintained via update triggers and decay factors. 
Updates can be executed in three options. Event-driven updates trust values quickly upon an event, such as a packet exchange, service invocation, or incoming recommendation. For instance, \cite{Latif2022A} uses event-driven updates to maintain trust ratings after every client-server communication.
Time-driven updates refresh scores at predetermined intervals, which is straightforward to manage. However, it may result in stale trust between updates. Hybrid methods integrate both techniques, updating on interaction when evidence is available and periodically updating otherwise \cite{Li2024Blockchain, Ahmad2025Adaptive}. Although frequently overlooked, trust decay is essential for history-based frameworks, as maintenance remains fundamentally incomplete without a temporal aging mechanism: fresh occurrences update the score, but old evidence may still dominate the outcome. Decay mechanisms discount older data to prioritize current conditions over past history. This is important in IoT because devices may become inactive, environments change rapidly, and attackers may exploit outdated high trust through on-off behavior. Decay is often implemented using forgetting factors such as sliding time windows, or separate trust terms. \cite{Sharma2025Evaluation} includes a recent trust in their suggestion, whereas \cite{Kumar2025AI} uses historical trust, where the final trust value is based on the current trust and the average of previous trust values.

\subsection{Trust-Related Attacks}\label{sec-foundations-attacks}

In the literature, several taxonomies have been proposed to categorize trust-related attacks. \cite{tyagi2023detailed} employs a basic taxonomy: internal, external attacks and their intersection. In addition to classification based on attacker location, \cite{hossain2024holistic} suggests a taxonomy of three more classes:  device specification, information damage level, and host compromise. Both \cite{sagar2024understanding} and \cite{marche2020trust, Moeinaddini2025A} use two-dimensional classifications: the former categorizes attacks as individual or collusive, while the latter groups them by target (service/recommendation) and impact scope (single/group). 

\begin{table*}[ht]
\centering
\footnotesize
\caption{Classification of Trust-Related Attacks by Target and Architectural IoT Layer}
\label{tab:attack_taxonomy_simplified}
\begin{tabular}{c p{2.7cm} p{3.2cm} p{1.2cm}}
\toprule
\textbf{Target} & \textbf{IoT Layer} & \textbf{Attack Type} & \textbf{Abbr.}  \\ 
\midrule

\multirow{3}{*}{\rotatebox{90}{\makecell{Device \\ Trust}}} 
 & Network/Application & Sybil Attack & SA  \\ 
 & Physical & Node Capture & NC  \\
 & Physical & Sleep Deprivation & SDA  \\
\addlinespace \hline \addlinespace

\multirow{7}{*}{\rotatebox{90}{\makecell{Data \\ Trust}}} 
 & Application & Bad-mouthing & BMA  \\
 & Application & Good-mouthing (Ballot-stuffing) & GMA  \\
 & Application & Self-promoting & SPA  \\
 & Application & Data Poisoning & DP  \\ 
 & Physical/Network & False Data Injection & FDI  \\
 & Physical/Network & Jamming & JA  \\
 & Physical & Sensor Spoofing & SS  \\

\addlinespace \hline \addlinespace

\multirow{8}{*}{\rotatebox{90}{\makecell{Service \\ Trust}}} 
 & Application & Opportunistic Service & OSA  \\  
 & Application & Discriminatory Attack  & DA  \\ 
 & Application &  Whitewash Attack  & WA  \\ 
 & Application &  Malicious with Everyone Attack  & ME  \\   
 & Network/Application & On-Off Attack & OOA  \\ 
 & Network & Selective Forwarding & SFA  \\
 & Network & Black-hole & BHA  \\
 & Network & Gray-hole & GHA  \\

\bottomrule
\end{tabular}
\end{table*}

Moreover, \cite{Yang2025PRADA} focuses specifically on data trust-related attacks and proposes a taxonomy based on three dimensions: attack methodology, attacker type (non-strategic vs. strategic), and attack scope (local vs. network-wide). While their approach may provide understanding of adversarial behavior, it only addresses trustworthiness of information within navigation systems. 
Finally, while \cite{konsta2022trust} maps trust attacks strictly to single IoT layers, our proposed taxonomy allows for hybrid layer assignments. As shown in Table \ref{tab:attack_taxonomy_simplified}, we classify each threat by its architectural layer and specific trust target (device, data, or service). Furthermore, Table~\ref{tab:attack-taxonomy-further} maps the same attacks onto the process in Figure ~\ref{fig:trust-management-process}. It records where the compromise occurs, its primary effect, whether the attack is internal or external, the behavior of attacker, and how difficult the attack is to detect.

\begin{table*}[ht]
\centering
\footnotesize
\caption{Trust-Related Attacks Across The Trust Management Process in Figure ~\ref{fig:trust-management-process}}
\label{tab:attack-taxonomy-further}
\begin{tabular}{p{0.44cm} p{3.94cm} p{5.8cm} p{0.77cm} p{1.014cm} p{1.071cm} }
\toprule
\textbf{Abbr.} &
\textbf{Targeted Trust Management Process} &
\textbf{Primary Effect} &
\textbf{Attack Origin} &
\textbf{Attacker Behavior} &
\textbf{Detection Difficulty} \\
\midrule
SA  & Evidence flow; Trust flow; Maintenance & Fake identities bias evidence and disseminated scores & I \& E & Persistent & $\blacksquare \blacksquare \blacksquare$\\ 
NC  & Evidence flow & A captured node becomes a false evidence source & I \& E & Persistent & $\blacksquare \blacksquare \square$ \\ 
SDA & Evidence flow & Honest nodes are exhausted and stop reporting & I \& E & Persistent & $\blacksquare \blacksquare \square$ \\
\addlinespace \hline \addlinespace
BMA & Acquisition & False recommendations slander honest nodes & I & Persistent & $\blacksquare \blacksquare \square$ \\
GMA & Acquisition; Trust flow & Colluding nodes inflate shared reputation & I & Strategic &  $\blacksquare \blacksquare \blacksquare$ \\
SPA & Acquisition; Trust flow & A node inflates the score that others receive about it & I & Persistent & $\blacksquare \blacksquare \square$ \\
DP  & Evidence flow; Computation & Corrupted samples bias the trust model itself & I \& E & Persistent &  $\blacksquare \blacksquare \blacksquare$ \\
FDI & Evidence flow; Aggregation & False observations enter the evidence stream & I \& E & Persistent & $\blacksquare \blacksquare \square$ \\
JA  & Evidence flow; Trust flow & Evidence collection and score dissemination are disrupted & E & Persistent & $\blacksquare \square \square$ \\
SS  & Evidence flow & Fabricated sensor readings replace genuine measurements & E & Persistent & $\blacksquare \blacksquare \square$ \\
\addlinespace \hline \addlinespace
OSA & Trust-driven decision flow & Service is degraded after the node has been selected & I & Strategic &  $\blacksquare \blacksquare \blacksquare$ \\
DA  & Trust-driven decision flow & Only selected requesters receive poor service & I & Intermittent &  $\blacksquare \blacksquare \blacksquare$ \\
WA  & Maintenance & Identity reset discards prior distrust & I & Strategic &  $\blacksquare \blacksquare \blacksquare$ \\
ME  & Trust-driven decision flow & Every trust-based request is served maliciously & I & Persistent & $\blacksquare \square \square$ \\
OOA & Trust-driven decision flow; Maintenance & Alternating behaviour keeps the node selectable & I & Intermittent &  $\blacksquare \blacksquare \blacksquare$ \\
SFA & Trust-driven decision flow & Trusted forwarding is applied only to some packets & I & Persistent & $\blacksquare \blacksquare \square$ \\
BHA & Trust-driven decision flow & Trusted routing delivers traffic that is then dropped & I & Persistent & $\blacksquare \square \square$ \\
GHA & Trust-driven decision flow & Trusted routing is violated only part of the time & I & Intermittent &  $\blacksquare \blacksquare \blacksquare$ \\
\bottomrule
\end{tabular}

\footnotesize{I = Internal Attack, E = External Attack, I \& E = Both Applicable}

\end{table*}

Table~\ref{tab:attack-taxonomy-further} reveals that trust-related attacks can compromise the trust management process at multiple stages rather than targeting trust computation alone. Evidence-related attacks such as SS, FDI, and BMA primarily corrupt the information used to establish trust, whereas OSA, DA, SFA, BHA, and GHA exploit trust-based decisions after a node has already been considered trustworthy. Attacks such as SA, DP, and OOA are particularly broad in their impact, spanning multiple stages of the process and potentially influencing both trust assessment and subsequent decisions. There is also a correlation between attacker behavior and detection difficulty. Persistent attacks generally leave continuous behavioral evidence and are therefore comparatively easier to detect, while intermittent and strategic attacks, including OOA, GHA, DA, GMA, OSA, and WA, deliberately preserve or recover favorable trust levels and are consequently harder to identify. This is especially evident for attacks that manipulate trust without continuously violating expected behavior. Overall, this representation shows that attack resilience should be evaluated at the process level, since a trust mechanism may accurately compute trust from its inputs yet remain vulnerable when those inputs, trust-based decisions, or subsequent trust updates are manipulated.

\section{Review Methodology} \label{sec-methodology}

\begin{figure*}[htbp]
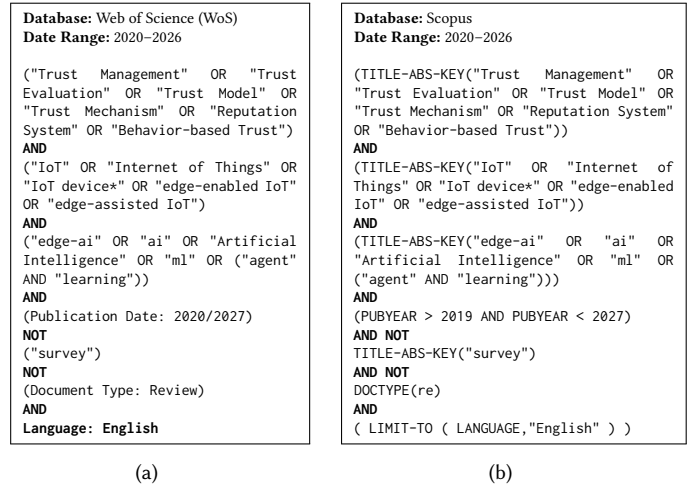

    \centering
    \scriptsize
\noindent
    \begin{minipage}{0.24\textwidth}
    \noindent
        \fbox{
            \parbox{\linewidth}{
                \textbf{Database:} Web of Science (WoS) \\
                \textbf{Date Range:} 2020--2026 \\ 
                \hrulefill \\
                \texttt{
                ("Trust Management" OR "Trust Evaluation" OR "Trust Model" OR "Trust Mechanism" OR "Reputation System" OR "Behavior-based Trust") 
                \\ \textbf{AND} \\
                ("IoT" OR "Internet of Things" OR "IoT device*" OR "edge-enabled IoT" OR "edge-assisted IoT") 
                \\ \textbf{AND} \\
                ("edge-ai" OR "ai" OR "Artificial Intelligence" OR "ml" OR ("agent" AND "learning")) 
                \\ 
                \textbf{AND} \\ (Publication Date: 2020/2027)  \\
                \textbf{NOT} \\
                ("survey") 
                \\ \textbf{NOT} \\
                (Document Type: Review)
                \\ \textbf{AND} \\
                \textbf{Language: English}
                }
            }
        }
        \subcaption{}
    \end{minipage}%
   \hspace{0.7cm} 
    \begin{minipage}{0.28\textwidth}
        \centering
        \fbox{
            \parbox{\linewidth}{
                \textbf{Database:} Scopus \\
                \textbf{Date Range:} 2020--2026 \\
                \hrulefill \\
                \texttt{
                (TITLE-ABS-KEY("Trust Management" OR "Trust Evaluation" OR "Trust Model" OR "Trust Mechanism" OR "Reputation System" OR "Behavior-based Trust")) 
                \\ \textbf{AND} \\
                (TITLE-ABS-KEY("IoT" OR "Internet of Things" OR "IoT device*" OR "edge-enabled IoT" OR "edge-assisted IoT"))
                \\ \textbf{AND} \\
                (TITLE-ABS-KEY("edge-ai" OR "ai" OR "Artificial Intelligence" OR "ml" OR ("agent" AND "learning")))
                \\ \textbf{AND} \\
                (PUBYEAR > 2019 AND PUBYEAR < 2027) 
                 \\ \textbf{AND NOT} \\
                TITLE-ABS-KEY("survey") 
                \\ \textbf{AND NOT} \\
                DOCTYPE(re) 
                \\ \textbf{AND} \\
                ( LIMIT-TO ( LANGUAGE,"English" ) )
                }
            }
        }
          \subcaption{}
    \end{minipage}
     \caption{Search Queries Applied to the Two Indexing Databases: (a) Web of Science (WoS) , and (b) Scopus}
    \label{fig:combined_search_queries}
    \Description{Search Queries}

\end{figure*}

To provide a comprehensive assessment of the current state of trust management in IoT, we adopt a systematic literature review approach. The primary repositories selected for this study are Web of Science (WoS) and Scopus. Constructed search queries given in Figure \ref{fig:combined_search_queries}.

\begin{figure}[h]
    \centering
    \includegraphics[width=0.85\linewidth]{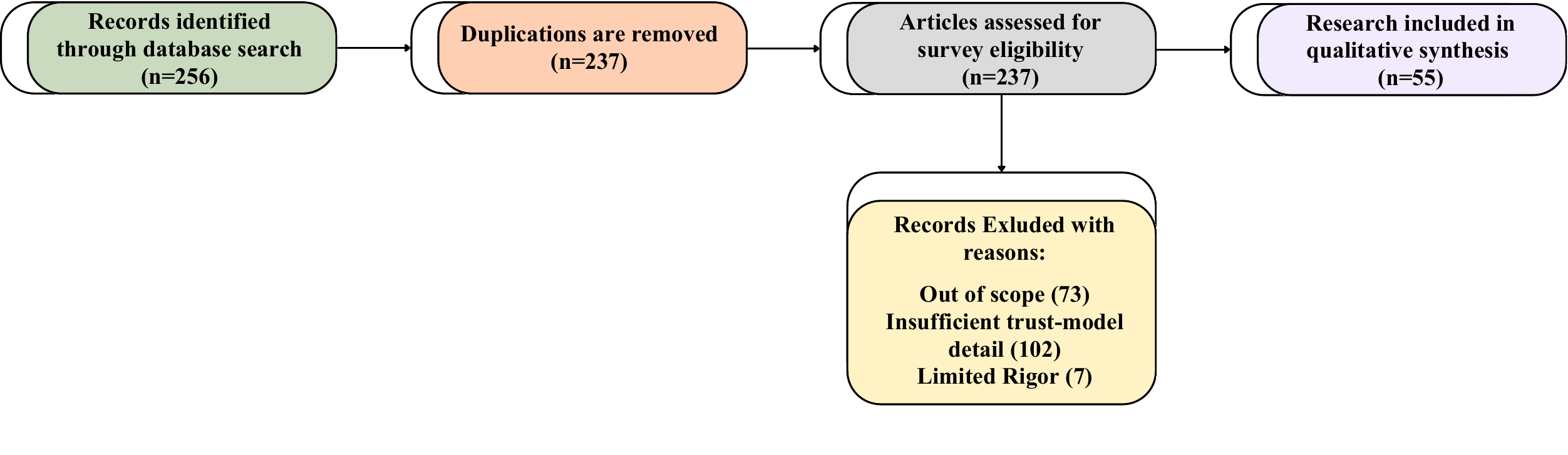}
    \Description{PRISMA Flow Diagram}
    \caption{PRISMA Flow Diagram From Identification To Screening, Eligibility, And Inclusion Steps}
    \label{fig:PRISMA}
\end{figure}

The review technique follows PRISMA framework to promote reproducibility and transparency in the article selection process. Figure \ref{fig:PRISMA} illustrates the steps involved in the literature screening and selection process.
Research included in qualitative synthesis are examined in steps \emph{(i) scope identification}, \emph{(ii) trust-design evaluation}, and \emph{(iii) operational and attack coverage}. 

\paragraph{Step 1: Scope identification}

Each state-of-the-art study is first assigned to one of the four IoT application domains considered in this survey: Consumer, Commercial, Industrial, or Infrastructure IoT. This classification is based on the primary operational context and type of IoT system investigated in the study. The study is then positioned according to the IoT layer at which trust is primarily computed, updated, or enforced, using the three-layer model of physical, network, and application layers. When trust mechanisms span multiple layers, a hybrid placement is recorded. This two-level classification establishes both the application context and architectural location of the trust mechanism.

\paragraph{Step 2: Trust-design evaluation}

\begin{figure}[h]
    \centering
    \includegraphics[width=0.95\linewidth]{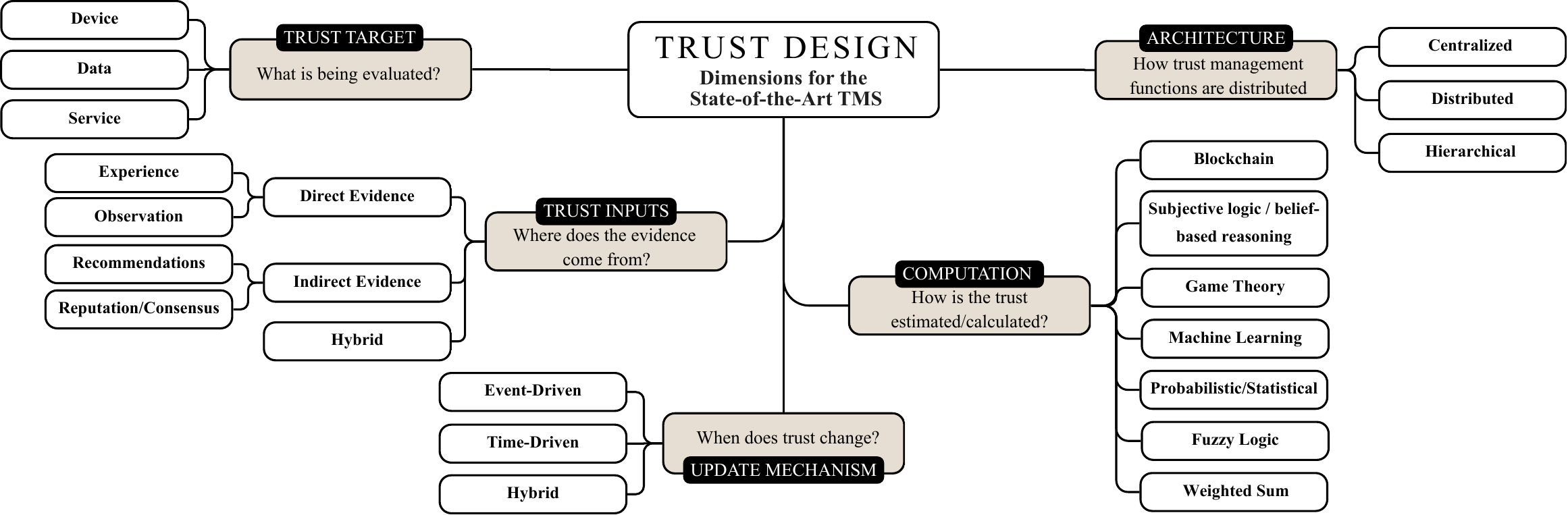}
    \Description{Trust Design Taxonomy}
    \caption{Trust-Design Taxonomy for Classifying Trust Management Schemes}
    \label{fig:trust-design-dimensions}
\end{figure}

Each study is further classified according to the five orthogonal trust-design dimensions presented in Figure~\ref{fig:trust-design-dimensions}. These dimensions capture what is trusted, how trust is computed, which evidence is used, how trust is updated or maintained. This stage enables the comparison of the underlying design choices.

\paragraph{Step 3: Operational and Attack Coverage}

The operational roles of trust are examined and results are given in Table~\ref{tab:trust_decisions}. For each study, we record how trust is used to securely guide IoT operations, such as routing, service selection, access control, data aggregation, or intrusion detection. The attacks reported as mitigated by each study are then extracted from the original paper and categorized according to the attack taxonomy in Table~\ref{tab:attacktable}.

\section{Current IoT Trust Based Frameworks} \label{sec-current-frameworks}

\begin{figure*}[h]
    \centering
    \begin{subfigure}{0.34\linewidth}
        \centering
        \includegraphics[width=\linewidth]{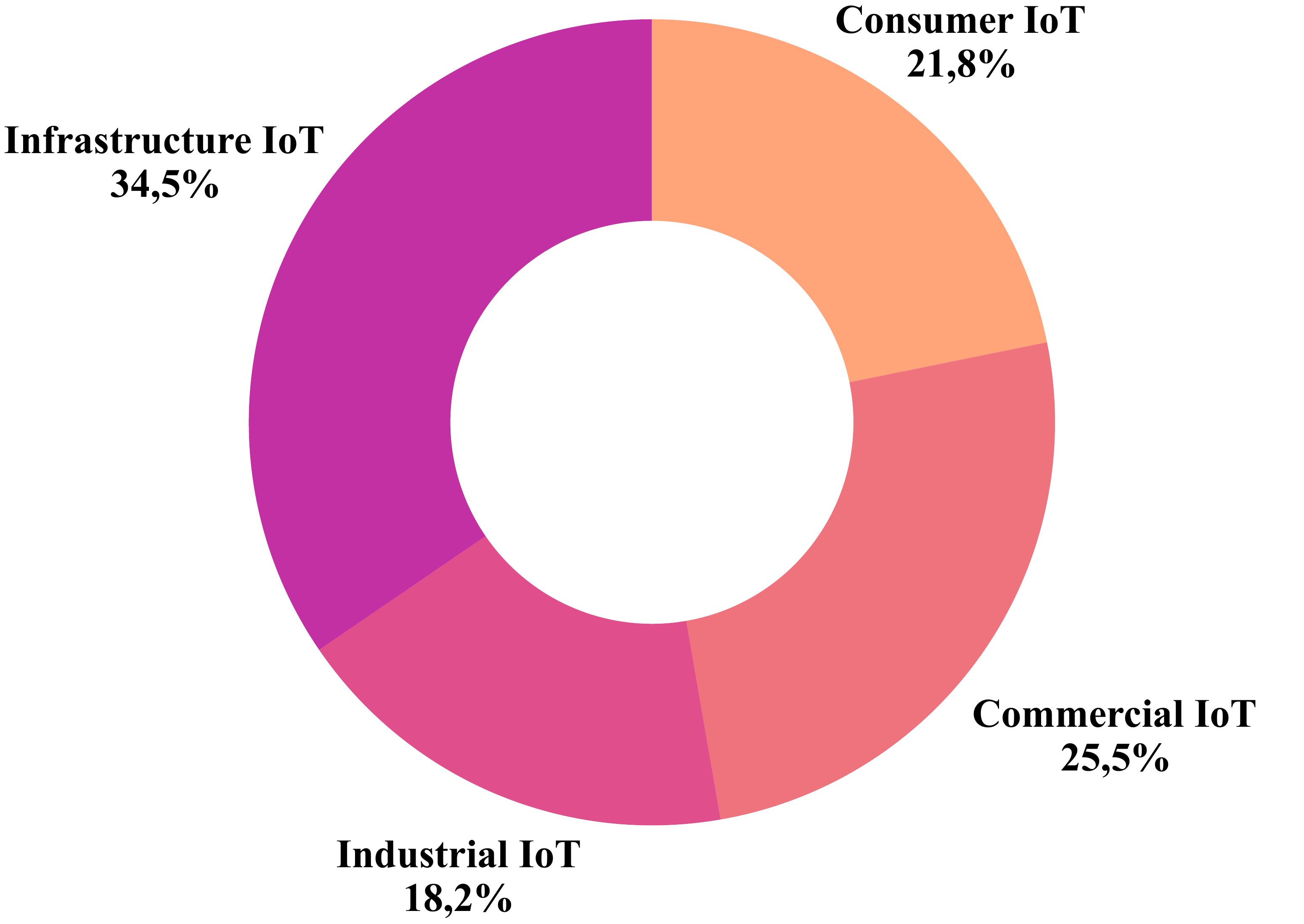}
        \Description{Distribution Across IoT Application Domains}
        \caption{}
        \label{fig:resultsub1}
    \end{subfigure}
   \hspace{1cm} 
    \begin{subfigure}{0.4\linewidth}
        \centering
        \includegraphics[width=\linewidth]{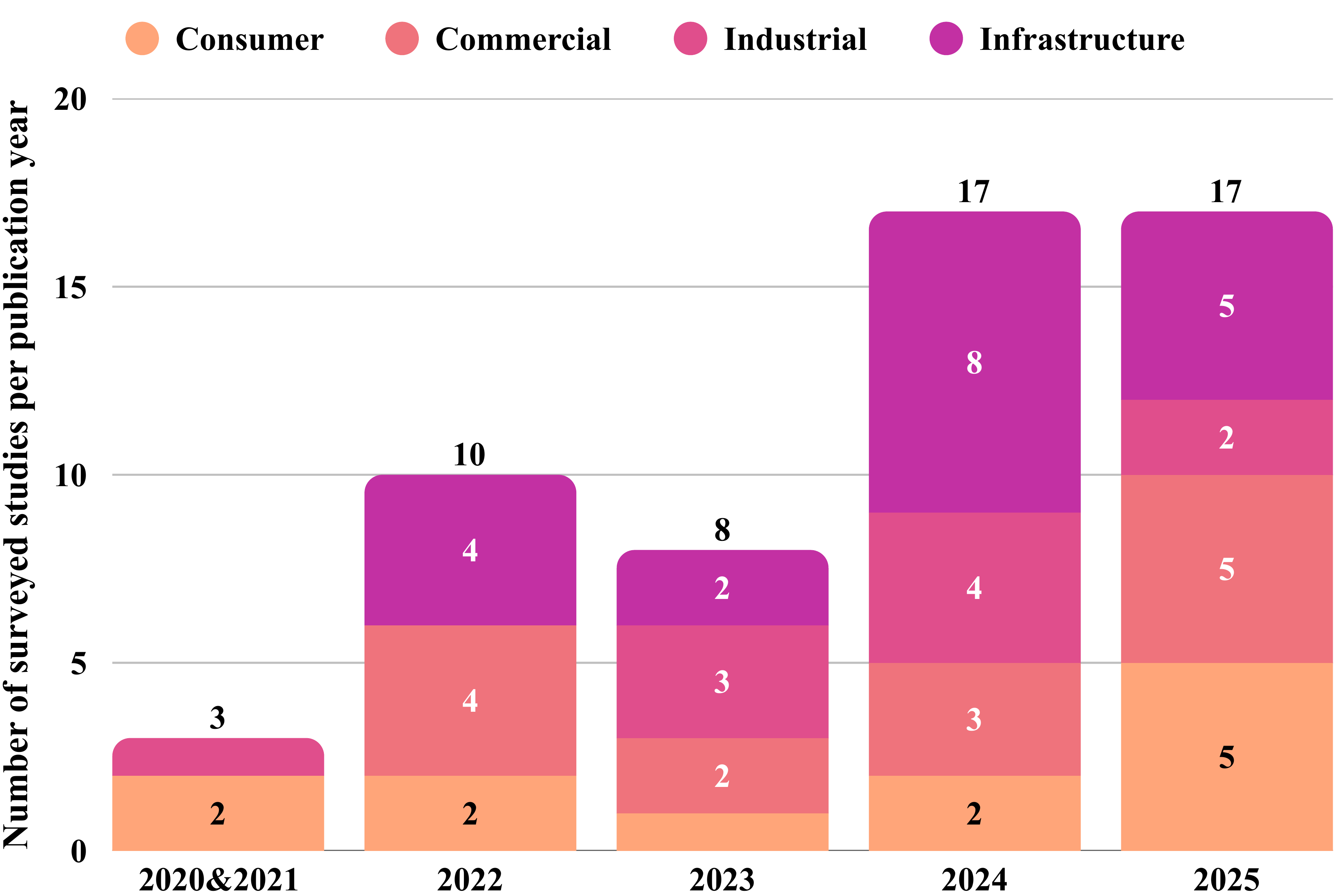}
        \caption{}
        \label{fig:resultsub2}
    \end{subfigure}
    \caption{Surveyed Literature Overview: (a) Distribution Across IoT Application Domains, and (b) Annual Distribution By Domain From 2020 to 2025}
    \label{fig:ResultSurvey}
\end{figure*}

We examine state-of-the-art trust management frameworks using the four-step classification from Section~\ref{sec-methodology}. The comparative taxonomies are organized by IoT application domain (consumer, commercial, industrial, infrastructure) and ordered top-down by operational layer (application to physical). Cross-layer mechanisms are explicitly indicated. Notably, most surveyed literature targets the application and network layers, where node cooperation and interaction primarily occur. The distribution respect to application domains is balanced across the four domains, although some years have limited coverage in specific domains as given in Figure \ref{fig:ResultSurvey}.

\begin{table*}[h] 
\centering
\caption{Comparative Taxonomy of Trust Management in Consumer IoT} 
\label{tab:taxonomy_comparison_consumer1}

\resizebox{\textwidth}{!}{%
	
\begin{tabular}{p{0.8cm} p{1.5cm} p{1.2cm} p{2.9cm} p{3.6cm} p{0.8cm} p{2.6cm} p{3cm} p{1.2cm} }
\toprule

\multicolumn{2}{c}{\textbf{Research Scope}} & 
\multicolumn{4}{c}{\textbf{Trust Design}} & 
\multicolumn{3}{c}{\textbf{Evaluation \& Security}}\\
\cmidrule(r){1-2} \cmidrule(lr){3-6} \cmidrule(lr){7-9} 

\textbf{Ref. (Year)} & 
\textbf{Evaluation Layer} & 
\textbf{Trust Target} &
\textbf{Computation Model} & 
\textbf{Trust Inputs} & 
\textbf{Update} &
\textbf{Attacks Mitigated} & 
\textbf{Validation} & 
\textbf{Code?}\\ 
\midrule


\cite{Fortino2023A} (2022) &
Application &
Device \& Service &
Weighted feedback aggregation, clustering, multi-agent &  
Peer feedback, resource relevance, interaction frequency-recency, trustor-prior reputation &
E &
Collusion, GMA, SPA, BMA, OOA, OSA &
Simulation & 
No \\

\hline

\cite{Mahmood2025Application} (2025) &
Application &
Device &
Logarithmic trust function, time-gated update &
Application transactions, authentication, timing gap, recommendations &
E \& T &
GMA, BMA, DoS/DDoS, flooding, SA, replay, JA, cloning &
Simulation on Cooja, vs LightTrust, ETES &
No 
\\
\hline
\cite{Moeinaddini2025A} (2025) &
Application &
Service &
MLP, FL &
Direct, recommendations &
E &
OOA, WA, OSA, BMA, GMA, SPA, DA, ME &
Simulation on Python &
No \\
\hline

\cite{Rehman2025A} (2025) & 
Application &
Device &
Hybrid fuzzy logic, FL  &
Sensor data, user behavior, context, historical interactions, real-time feedback &
E \& T &
OOA, WA, DDoS, GMA, BMA &
Simulation on MATLAB, vs MetaCIDS, MetaverseAuth, TrustITS &
Available upon request 
\\
\hline


\hline
\cite{Kuppusamy2024IoT} (2024) &
Application \& Network &
Device \& Data &
Direct, recommended, belief-based trust with threshold classification, clustering, RNN classifier &

Direct one-hop behavior, recommendations, belief trust, reputation from network features &

T &
External attacks, double-spending, collusion & 
Experimental evaluation on Python & 
Available upon request \\

\hline
\cite{Le2022Artificial} (2022) &
Application \& Network &
Device \& Data &
GAN, VAE, FL, blockchain consensus, reputation/data-similarity trust &
Historical interactions, neighbor opinions, data similarity, signal strength, voting, social trust, computation contribution &

-- &
SA, DP, FDI, eavesdropping, OSA &
Conceptual, healthcare HAR use-case  & 
No \\

\hline
\cite{Zhang2021AIT} (2021) &
Application \& Network &
Device \& Data &
Feedforward NN, blockchain (Merkle-tree trust archiving) &

18 NN inputs (vehicle/incident location, visibility distance, current trust, RSU range, traffic direction), message context, LTL/GTL trust &

E \& T &
SA, BMA, OOA, fake/tampered messages, compromised RSU &
Simulation on SUMO, NS-2 & 
No \\


\hline
\cite{Alsheakh2020Towards} (2020) &
Network &
Device &
Explainable Bayesian trust scoring with severity-aware weighting &

Service-access matches, access violations, access uncertainty, cumulative uncertain volume, access violation diversity, MUD/ACL baselines &
T &
DDoS, reflection attacks, flooding, malware &
Experimentation on real PCAP datasets (UNSW, CTU), SVM threshold classification &
No \\
\hline
\cite{Bampatsikos2025Trust} (2025) &
Network &
Device &
2D Markov chain, weighted-sum trust score, AHP/TOPSIS & 
CPU/RAM usage, security level (TEE/PUF RoT), OS/package age, packet loss, risk/reputation, vulnerability index, IDS score & 
E &
DDoS, NC, FDI, GMA, BMA, SPA, APTs &
Real-time testbed & 
No \\
\hline
\cite{Moudoud2022Detection} (2022) &
Network &
Device \& Data &
HMM behavior prediction, stake-based voting game&
HMM global trust, neighbor recommendations, past interactions, stake/deposit, vote validity, data skewness &

E \& T &
FDI, single-point-of-failure &
Simulation on Python, FDI dataset; compared with LSTM, GRU, RSS, Sec5G &
No \\
\hline

\cite{Ali2024TIHCS} (2024) &
Network \& Physical &
Device \& Data &
Weighted direct and indirect trust with sigmoid function, threshold classification &
GTS-request patterns, received vs requested packets, channel capacity, neighbor trust &
E \& T &
Slot-capturing DoS, impersonation, FDI &
Simulation, vs FCFS, RR, SJF, LJF, & 
No \\
\hline

\cite{Ahmed2025Trust} (2025) &
Physical &
Device &
FL, feed-forward ANN, Z-score/MAD outlier detection &
RSSI, LQI, internal temperature, battery level, MAC, radio channel, antenna orientation &
T &
SS, replay, SA, DoS &
Zigbee Z1 testbed dataset (347{,}200 instances) &
No \\
\bottomrule
\end{tabular}%
}
\footnotesize{E = Event-driven Update, T = Time-driven Update, E \& T = Hybrid, - = Not Reported}

\end{table*} 

To start with, Table \ref{tab:taxonomy_comparison_consumer1}
lists the investigated IoT trust frameworks in consumer IoT application domain. 
The main goal in consumer IoT is to make everyday connected systems more dependable without sacrificing usability in general. In this domain, a large share of this research focuses primarily on device trust \cite{Mahmood2025Application, Rehman2025A, Alsheakh2020Towards, Bampatsikos2025Trust, Ahmed2025Trust}, while service trust and data trust receive comparatively less attention. A second group assess trustworthiness of data together with device trust \cite{Kuppusamy2024IoT, Le2022Artificial, Ali2024TIHCS, Moudoud2022Detection, Zhang2021AIT}. Only \cite{Fortino2023A} jointly considers service trust and device trust, and \cite{Moeinaddini2025A} is the only research we surveyed that addresses service trust alone in this application area. 
The choice of trust inputs closely follows the usage of which trust target in evaluation layer. Physical-layer schemes such as~\cite{Ahmed2025Trust} use link and device-state signals and therefore focus on device trust. Network-layer works including~\cite{Alsheakh2020Towards, Bampatsikos2025Trust} assess device trust from communication and operational evidence, while~\cite{Moudoud2022Detection, Ali2024TIHCS} extend this to data trust when integrity-related inputs such as data skewness or received-versus-requested packets are added. Application-layer studies such as~\cite{Mahmood2025Application, Rehman2025A} mainly target device trust using transactions, authentication, and contextual analytics;~\cite{Fortino2023A} and~\cite{Moeinaddini2025A} are the main exceptions that explicitly consider service trust. Cross-layer schemes \cite{Kuppusamy2024IoT, Le2022Artificial, Zhang2021AIT}  evaluate device and data trust simultaneously. \cite{Kuppusamy2024IoT, Le2022Artificial} are both in the healthcare field, while \cite{Zhang2021AIT} uses trust management specifically for vehicular networks.
\cite{Kuppusamy2024IoT} recommends using direct one-hop behavior, neighbor recommendations, and belief-based reputation from network features as trust inputs. Furthermore, these are aggregated with threshold rules and clustering before trust gates access to electronic health records. Similarly, \cite{Le2022Artificial} uses neighbor opinions, signal strength, voting, and data-similarity signals as inputs, then combines FL, GAN/VAE sanitization, and blockchain-supported reputation to judge health-activity data trust. In \cite{Zhang2021AIT}, authors feed location, mobility, RSU range, message context, and prior local/global trust inputs into a feedforward neural network, while blockchain stores updated trust values for later validation of traffic and incident messages.

Computation models in consumer IoT fall into three overlapping groups. The largest group applies weighted aggregation of direct and indirect evidence \cite{Fortino2023A, Mahmood2025Application, Kuppusamy2024IoT, Bampatsikos2025Trust, Ali2024TIHCS}. A second group uses probabilistic or statistical inference \cite{Alsheakh2020Towards, Bampatsikos2025Trust, Moudoud2022Detection}, and a third learns the mapping from behavior to trust \cite{Kuppusamy2024IoT, Moeinaddini2025A, Rehman2025A, Le2022Artificial, Zhang2021AIT, Ahmed2025Trust}. The groups differ mainly in how the score is shaped and what it estimates. Within weighted aggregation, \cite{Mahmood2025Application} applies a logarithmic function so that trust accumulates slowly, whereas \cite{Ali2024TIHCS} applies a logistic function that sharpens the decision boundary, and \cite{Fortino2023A} places the security property in the weights themselves by scaling feedback with resource relevance, interaction frequency, and trustor reputation to resist collusion. Weights are otherwise hand-tuned, and \cite{Bampatsikos2025Trust} is the only study deriving them formally through AHP and TOPSIS. In the probabilistic group, \cite{Alsheakh2020Towards} scores past behavior through Bayesian updating with severity-aware discounting, while \cite{Bampatsikos2025Trust, Moudoud2022Detection} predict future behavior through Markov and hidden Markov models, which is valuable where interaction history is short and sparse. Where learning is used, the architectures are typically small feed-forward networks over low-dimensional feature vectors \cite{Moeinaddini2025A, Zhang2021AIT, Ahmed2025Trust}, whereas deeper generative and recurrent models appear only in \cite{Le2022Artificial, Kuppusamy2024IoT}. Only \cite{Rehman2025A} adds fuzzy inference with fractional-order dynamics for stability guarantees, and only \cite{Moudoud2022Detection} makes honest participation explicit through stake-based voting, while subjective logic and evidence-theoretic fusion are absent.
Earlier works rely on Bayesian scoring, neural inference, and blockchain reputation, whereas federated designs primarily in 2025. 
Federated and edge-local training is therefore the most recent trend in this domain, consistent with the privacy sensitivity of consumer-related evidence. 
Among the studies in Table \ref{tab:taxonomy_comparison_consumer1}, BMA \cite{Fortino2023A, Mahmood2025Application, Moeinaddini2025A, Rehman2025A, Zhang2021AIT, Bampatsikos2025Trust} and GMA \cite{Fortino2023A, Mahmood2025Application, Moeinaddini2025A, Rehman2025A, Bampatsikos2025Trust} are the most common, together with OOA \cite{Fortino2023A, Moeinaddini2025A, Rehman2025A, Zhang2021AIT}. Among conventional threats, SA \cite{Mahmood2025Application, Le2022Artificial, Zhang2021AIT, Ahmed2025Trust}, DDoS \cite{Mahmood2025Application, Rehman2025A, Alsheakh2020Towards, Bampatsikos2025Trust}, and FDI \cite{Le2022Artificial, Bampatsikos2025Trust, Moudoud2022Detection, Ali2024TIHCS} receive the most attention, while other attack classes appear only rarely. Routing attacks such as blackhole, grayhole, and selective forwarding are largely absent from Consumer IoT studies, indicating that they are treated as more relevant to multi-hop networking domains such as Industrial IoT or Infrastructure IoT. 

\begin{table*}[h] 
\centering
\caption{Comparative Taxonomy of Trust Management in Commercial IoT}
\label{tab:taxonomy_comparison-commercial1}
\resizebox{\textwidth}{!}{%
	
\begin{tabular}{p{0.8cm} p{1.5cm} p{1.2cm} p{2.9cm} p{3.6cm} p{0.8cm} p{2.6cm} p{3cm} p{1.2cm} }
\toprule

\multicolumn{2}{c}{\textbf{Research Scope}} & 
\multicolumn{4}{c}{\textbf{Trust Design}} & 
\multicolumn{3}{c}{\textbf{Evaluation \& Security}}\\
\cmidrule(r){1-2} \cmidrule(lr){3-6} \cmidrule(lr){7-9} 

\textbf{Ref. (Year)} & 
\textbf{Evaluation Layer} & 
\textbf{Trust Target} &
\textbf{Computation Model} & 
\textbf{Trust Inputs} & 
\textbf{Update} &
\textbf{Attacks Mitigated} & 
\textbf{Validation} & 
\textbf{Code?}  
\\ 
\midrule


\cite{Ahmad2025Adaptive} (2025) &
Application &
Device &
Behavior-based trust &
Reputation & 
T & 
SA, OOA, replay, DoS, MiTM &
Implementation on Raspberry Pi, ESP32 &
No \\
\hline
\cite{Rathee2025An} (2025) &
Application &
Device &
4-dimensional trust mechanism, blockchain surveillance &
Sensor behavior, energy consumption, forwarding behavior, activeness, delay, information accuracy &

T &
Abnormal device behavior, control-command, coordinated threats &

Simulation on MATLAB &
Available upon request \\

\hline
\cite{Wang2025A} (2025) &
Application &
Service &
Weighted trust evaluation &
Hybrid &
E & 
BMA, GMA, DA &
Simulation &
No \\

\hline
\cite{Sun2024Trust} (2024) &
Application &
Service &
Interval MADM trust evaluation, deviation-maximization weighting &

QoS attributes via SLO, interval-valued monitoring values &

E &
Not reported &
Case study on QWS dataset  & 
No \\
\hline
\cite{Latif2022A} (2022) &
Application &
Device \& Service &
MCDA (QoS-weighted ratings), incremental SVD collaborative filtering &

QoS params, peer post-communication ratings &
E &
DoS, DDoS, malicious edge nodes, FDI &
Simulation on EdgeCloudSim, MATLAB & 
No \\
\hline
\cite{Liang2024Collaborative} (2024) &
Application &
Service &
Blockchain-based reputation, short-term incentives, HRL service optimization &


Service-provision records, service collaboration frequency, direct/indirect reputation, rewards/punishments, edge resources &
E \& T &
OSA, unreliable/malicious SP &
Simulation on Python, PyTorch, train-control case study & 
No \\
\hline

\cite{Tsekenis2025Flexible} (2025) &
Network &
Device &

Weighted sum, ML-enhanced genetic algorithm&
Latency, throughput, capacity, energy efficiency, cost, reliability &
E \& T &
Not reported &
Simulation & 
No \\

\hline
\cite{Tuncel2025SAFE} (2025) &
Network &
Device &
Weighted multi-parameter trust on blockchain, FL K-means clustering, multi-agent RL &

Direct, indirect, and recent trust &
T &
Tampering, eavesdropping, privacy, single-point-of-failure, compromised nodes &
Simulation on NS-3, vs EDDC, LSTM, GRADE & 
Yes \\
\hline
\cite{Sharma2024Multi} (2024) &
Network &
Device \& Data &
Multi-level weighted trust aggregation &
Direct \& Indirect &
T &
BMA, GHA, FDI, replay, eavesdropping &
Simulation with LT-FS-ID dataset & 
No \\
\hline
\cite{Chen2023A} (2023) &
Network &
Device &
Reputation-record trust (request-count thresholding), multi-agent DRL & 

Resource request count, task-completion ability under delay constraint, channel state &

E &
Resource preemption, interference, JA, data tampering &
Simulation, Lee/Hata channels, vs KNN, A2C &

No \\
\hline
\cite{Yu2023CET} (2023) &
Network &
Device \& Data &
Neural network, fuzzy logic &
Direct/virtual interaction metrics, fuzzy attributes &
E &
FDI, malicious terminal detection &
Simulation on MATLAB with SmartSantander dataset &
No \\

\hline
\cite{Guo2022ITCN} (2022) &
Network &
Device \& Data &
Ground-truth matching, recommender-weighted indirect trust &
Direct \& Indirect &
E \& T &
FDI &
Simulation on Python with T-Drive Beijing taxi dataset & 
No \\
\hline
\cite{Guo2022Endogenous} (2022) &
Network &
Service &
Delay-based reputation, RPBFT, A3C DRL &
Service delay, historical reputation, QoS/praise &

E &
Byzantine/malicious nodes &
Simulation on Python, Hyperledger Fabric, vs PBFT, DQN, RD & 
No \\

\hline
\cite{Salim2022SEEDGT} (2022) &
Network \& Physical &
Device \& Data &
Weighted sum of behavioral metrics &
Packet/forwarding rates, energy, residual energy, distance, BS feedback &
T &
Sinkhole, internal/external attacks, known/chosen-plaintext attacks &
Simulation on MATLAB & 
No \\

\bottomrule
\end{tabular}
}
\footnotesize{E = Event-driven Update, T = Time-driven Update, E \& T = Hybrid, - = Not Reported}

\end{table*}

Secondly, Table \ref{tab:taxonomy_comparison-commercial1} lists the investigated IoT trust frameworks in the commercial IoT application domain.
Commercial IoT trust work mainly aims to keep enterprise services dependable: filter bad devices, select good services, and support efficient edge/cloud operation. Unlike consumer IoT, the emphasis is less on privacy and more on service reliability, scalability, and institutional trust in business environments. Device trust is still the most popular trust target in commercial IoT, as seen in \cite{Ahmad2025Adaptive, Rathee2025An, Tsekenis2025Flexible, Tuncel2025SAFE, Chen2023A, Latif2022A, Yu2023CET, Guo2022ITCN, Salim2022SEEDGT}. Similar attention is given to assessing trust considering both device and data \cite{Sharma2024Multi, Yu2023CET, Guo2022ITCN, Salim2022SEEDGT}. For example, \cite{Guo2022ITCN} computes data trust as the matching degree between a vehicle's reported data and UAV-collected ground-truth baselines, updating the score through bounded increase and decrease rules that converge as it approaches its limits. However, data trust is underexplored in this domain. 
Frameworks assessing service trust \cite{Wang2025A, Sun2024Trust, Liang2024Collaborative, Guo2022Endogenous, Latif2022A} are more visible than the number of frameworks in consumer IoT. When compared to consumer IoT, a greater attention is given to service trust. It is incorporated in 36\% of the surveyed commercial IoT studies, compared with only 17\% in Consumer IoT. In the research by \cite{Latif2022A}, QoS-driven peer ratings are aggregated into community trust and supports cold-start prediction for new devices. More, \cite{Guo2022Endogenous} applies delay-based service reputation to cross-domain service orchestration and consensus leader selection. 
Moreover, application layer is the primary deployment point for service trust mechanisms. For example, \cite{Sun2024Trust} ranks edge/cloud providers using QoS evidence such as availability, latency, and response time through a trust-as-a-service broker, while \cite{Wang2025A} evaluates edge service collaboration with multi-level trust, whitelist/blacklist control, and reputation attacks such as BMA and GMA. \cite{Liang2024Collaborative} further uses blockchain-based service reputation and collaboration records to schedule trustworthy edge services in train-control environments. In contrast, network layer more often evaluates device and data trustworthiness jointly. For example, \cite{Sharma2024Multi} considers interaction, data, and transmission trust between the members of a cluster, and identity and validation trust at the cluster-head and the base-station levels. They utilized direct observation across a sliding time window and, in the absence of interaction history, from neighbor input. The aggregated score is then used to drive cluster-head selection and routing. Overall, they try to mitigate GHA, OOA, WA, and BMA.
Besides \cite{Yu2023CET} evaluates terminal trust from interaction accuracy and response quality to detect FDI in smart-city sensing. In \cite{Salim2022SEEDGT}, forwarding behavior is linked with trusted data aggregation to prevent sinkhole attacks in cluster-based networks. \cite{Tsekenis2025Flexible} reframes trustworthiness as a multi-criteria operational metric for 6G mesh node selection using latency, cost, and capacity evidence, but it is not security-oriented and includes no attack model.
Consequently, trust inputs such as packet delivery rate \cite{Tuncel2025SAFE}
forwarding rate and packet success \cite{Salim2022SEEDGT}, latency, cost, and capacity \cite{Tsekenis2025Flexible}, 
channel gain and task completion under delay \cite{Chen2023A} are the trust inputs considered in these groups of frameworks. The trust model in \cite{Tuncel2025SAFE} combines direct trust derived from observed node behavior and packet delivery rate with indirect trust obtained from neighboring nodes' opinions, while recent trust values are further derived from the direct and indirect trust assessments.

\begin{table*}[h] 
\centering
\caption{Comparative Taxonomy of Trust Management in Industrial IoT}
\label{tab:taxonomy_comparison-industrial1}

\resizebox{\textwidth}{!}{%
	
\begin{tabular}{p{0.8cm} p{1.5cm} p{1.2cm} p{2.9cm} p{3.6cm} p{0.8cm} p{2.6cm} p{3cm} p{1.2cm} }
\toprule

\multicolumn{2}{c}{\textbf{Research Scope}} & 
\multicolumn{4}{c}{\textbf{Trust Design}} & 
\multicolumn{3}{c}{\textbf{Evaluation \& Security}}\\
\cmidrule(r){1-2} \cmidrule(lr){3-6} \cmidrule(lr){7-9} 

\textbf{Ref. (Year)} & 
\textbf{Evaluation Layer} & 
\textbf{Trust Target} &
\textbf{Computation Model} & 
\textbf{Trust Inputs} & 
\textbf{Update} &
\textbf{Attacks Mitigated} & 
\textbf{Validation} & 
\textbf{Code?} 
\\ 
\midrule


\cite{Ratnayake2024Machine} (2024) &
Application &
Data &
MLP, random forest, KNN &
Reputation score, SVM decision score & 
E & 
FDI, data manipulation &
Dataset-based experiments, prototype &
No \\
\hline
\cite{Wang2025Attribute} (2025) &
Application &
Device &
Weighted sum (direct + indirect) &
Decryption feedback, interaction outcomes-time-frequency &
E &
FDI &
Formal proofs, JPBC benchmarks, Python simulation with synthetic data &

No \\

\hline
\cite{Cheng2024Feedback} (2024) &
Application &
Device &
Feedback-based trust, Mahalanobis similarity, median filtering &
Reliability, communication success, node importance, interaction time, feedback &
E \& T &
BMA, OOA &
Simulation, vs TFL-DT, LTrust & 
No \\
\hline
\cite{Yu2024Dynamic} (2024) &
Application &
Device &
Feedback-based weighted trust (direct + indirect) &

Satisfaction, interaction history, time decay, operation counts, behavioral similarity &
T &
BMA, GMA, OOA, Collusion &
Simulation on NetLogo & 
No \\

\hline

\cite{Feng2023Blockchain} (2023) &
Application &
Device \& Service &
Blockchain-based trust, threshold, majority aggregation &
Prediction accuracy/timeliness, device feedback, running-state changes, periodic feedback  &
E \& T &
SPA, data tampering, false predictions, BMA &
Simulation &
Available upon request \\
\hline
\cite{Li2023Federated} (2023) &
Application &
Device &
FL with LSTM anomaly detection, weighted trust &
Consensus participation, behavior detection, traffic anomaly score, historical trust &
E \& T &
SA, impersonation, insider, collusion &
Experiments on Hyperledger Fabric/Docker, ZTE traffic &
No \\
\hline


\cite{Kumar2025AI} (2025) &
Network &
Device &
SVM, dynamic weighting &
Payload size, reputation, integrity, network statistics &
E &
Spoofing, SA, NC, replay, DDoS &
Prototype, Contiki-NG, synthetic dataset & 
Available upon request \\
\hline
\cite{Zhang2023A} (2023) &
Network &
Device \& Data &
Blockchain consensus trust, BLS-based PoRep via VDF &
Replica proofs, storage capacity, VDF verification, BLS signatures &

E &
Storage cheating, data tampering/deletion, TTP risk &

Experiments on C++/MIRACL library & 
Available upon request \\

\hline
\cite{Wang2024Secure} (2024) &
Network \& Physical &
Device &
PKI-based trusted authentication, behavior trust &
Identity/certificates, platform integrity, network behavior &
E \& T &
FDI, SFA, replay, eavesdropping, impersonation, MitM &
Simulation on NS-3 \& MATLAB & 
No \\
\hline
\cite{Wang2020MTES} (2020) &
Physical &
Device &
Probabilistic graphical model, approximation algorithm &
Behavior on Data-collection-communication, direct interactions & 
\-- &
Insider, hidden data, malicious/damaged nodes &
Simulation on MATLAB,NS-3 & 
No \\

\bottomrule
\end{tabular}
} 
\footnotesize{E = Event-driven Update, T = Time-driven Update, E \& T = Hybrid, - = Not Reported}

\end{table*}

With respect to computation, the commercial IoT literature can be summarized in three groups. The largest group derives trust from weighted, multiple attribute decision making (MADM), or rule-based aggregation of behavioral and QoS evidence \cite{Ahmad2025Adaptive, Rathee2025An, Wang2025A, Sun2024Trust, Sharma2024Multi, Salim2022SEEDGT}. A second group employs learning only to optimize a downstream decision, such as offloading, service orchestration, clustering, or topology formation \cite{Chen2023A, Guo2022Endogenous, Liang2024Collaborative, Tuncel2025SAFE, Tsekenis2025Flexible}. A third, much smaller group estimates trust itself with a learned model: \cite{Latif2022A} uses SVD for cold-start prediction, and \cite{Yu2023CET} combines a NN with fuzzy attributes. Where blockchain is used, it serves as a logging or consensus mechanism \cite{Rathee2025An, Liang2024Collaborative, Tuncel2025SAFE, Guo2022Endogenous}.

Three contrasts follow from this grouping. First, relative to consumer IoT, commercial studies give more attention to service-oriented MADM and reputation, yet none adopt a probabilistic or Bayesian computation. Second, within the domain, application-layer schemes typically rank services or providers, whereas network-layer schemes combine weighted device and data scores with an optimizer.
Third, in commercial IoT, event-driven trust updates are a common approach by 43\% \cite{Wang2025A, Sun2024Trust, Latif2022A, Chen2023A, Yu2023CET, Guo2022Endogenous}. In these frameworks, trust scores are revised after service collaboration, provider interaction, or QoS change. Time-driven mechanisms are also slightly more frequent in commercial IoT by 36\% particularly in schemes that require continuous monitoring of device behavior \cite{Ahmad2025Adaptive, Rathee2025An, Tuncel2025SAFE, Sharma2024Multi, Salim2022SEEDGT}.

Thirdly, frameworks we investigated that fall into industrial IoT domain are listed in Table \ref{tab:taxonomy_comparison-industrial1}. These frameworks mainly aim to preserve safe and continuous cyber-physical operation in factory, and sensor-network environments. Although the studies target different entities, they collectively seek to ensure that only trustworthy devices, data, edge services, and communication paths participate in industrial monitoring and control. Device trust is the dominant target in all three domains, but industrial IoT shows the strongest concentration on standalone device trust by 70\% in all layers. \cite{Wang2025Attribute, Cheng2024Feedback, Yu2024Dynamic, Li2023Federated, Kumar2025AI, Wang2024Secure, Wang2020MTES} assess device trustworthiness, in addition to that, \cite{ Zhang2023A} extends this by including data trustworthiness and \cite{Feng2023Blockchain} by integrating digital twins and blockchain, where trust scores guide autonomous service and server selection to support self-healing Edge-AI IIoT.
The only research focus solely on data trust is \cite{Ratnayake2024Machine} where it addresses a core limitation of blockchain-enabled IoT: blockchain ensures tamper-resistant storage, on the other hand, it does not guarantee that uploaded sensor data is accurate or trustworthy. The authors therefore aim to evaluate data trust at the point of submission by combining blockchain-based past device reputation with real-time SVM-assisted data classification at the edge, and validator-driven ML ensemble verification for uncertain records. 
Although this context dominated by device trust, FDI remains the most frequently addressed attack \cite{Ratnayake2024Machine, Wang2025Attribute, Feng2023Blockchain, Zhang2023A, Wang2024Secure}. It is because compromised industrial devices are seen as the main source of corrupted process data. Data quality and service-output validity are used as an additional evidences for device trustworthiness \cite{Wang2025Attribute, Feng2023Blockchain, Wang2024Secure}. Only \cite{Ratnayake2024Machine} accept data trust in its evaluation scope, whereas the remaining device-trust studies mainly address reputation \cite{Cheng2024Feedback, Yu2024Dynamic}, insider \cite{Wang2020MTES,Li2023Federated}, collusion \cite{Yu2024Dynamic, Li2023Federated}, or network attacks \cite{Kumar2025AI}. \cite{Yu2024Dynamic} jointly models device interaction and communication behavior to broaden attack coverage, but it still treats the edge broker as implicitly trusted; the authors therefore suggest future direction of a  trusted execution environment at the edge device.

Computationally, industrial IoT trust is less a single modeling tradition than a set of mechanisms aligned with process integrity. Weighted feedback remains the most common formulation, aggregating terminal and device interaction evidence at the edge \cite{Cheng2024Feedback, Yu2024Dynamic}. Commercial IoT uses similar aggregation, but typically to rank services or allocate resources through QoS and MCDA criteria \cite{Sun2024Trust, Latif2022A}. Industrial schemes instead filter MEC feedback and process data, treating trust as a gate on operational evidence rather than as a service-selection score. In learning-based models, Consumer IoT frameworks more often combines FL with fuzzy inference, generative sanitization, or probabilistic predictors for home and healthcare settings \cite{Rehman2025A, Le2022Artificial}. In industrial IoT, by contrast, it is used mainly for anomaly detection and rejection of corrupted sensor readings \cite{Ratnayake2024Machine, Li2023Federated}. Blockchain appears in all three domains, yet its role differs. In industrial deployments, it supports cross-domain federation, digital-twin validation, and storage-consensus trust \cite{Feng2023Blockchain, Zhang2023A}, rather than the reputation archiving typical of consumer IoT \cite{Zhang2021AIT} or the service-market logging typical of commercial IoT \cite{Liang2024Collaborative}. Trust updates in industrial IoT are mainly event-driven or hybrid (40\% each). Scores change when data are submitted, interactions occur, or consensus proofs complete. Effective maintenance is important not to leave old trust score in the loop.

\begin{table*}[h] 
\centering
\caption{Comparative Taxonomy of Trust Management in Infrastructure IoT}
\label{tab:taxonomy_comparison-infrastructure1}

\resizebox{\textwidth}{!}{%
	
\begin{tabular}{p{0.8cm} p{1.5cm} p{1.2cm} p{2.9cm} p{3.6cm} p{0.8cm} p{2.6cm} p{3cm} p{1.2cm} }
\toprule

\multicolumn{2}{c}{\textbf{Research Scope}} & 
\multicolumn{4}{c}{\textbf{Trust Design}} & 
\multicolumn{3}{c}{\textbf{Evaluation \& Security}}\\
\cmidrule(r){1-2} \cmidrule(lr){3-6} \cmidrule(lr){7-9} 

\textbf{Ref. (Year)} & 
\textbf{Evaluation Layer} & 
\textbf{Trust Target} &
\textbf{Computation Model} & 
\textbf{Trust Inputs} & 
\textbf{Update} &
\textbf{Attacks Mitigated} & 
\textbf{Validation} & 
\textbf{Code?}  
\\ 
\midrule


\cite{Din2025Building} (2025) &
Application &
Device &
FL, GenAI &
Vehicle behavior, communication patterns, telemetry, recommendations, data credibility &
E &
Insider &
Simulation on CityPulse dataset, real-world urban traffic &
No \\
\hline
\cite{Xiang2025Secure} (2025) & 
Application &
Data &
Weighted sum (direct+recommendation trust), DL anomaly detection &
Interaction satisfaction-success/failure, recommendations, data correctness &
E &
NC, FDI, OOA, BMA, impersonation, replay, MitM, ESL &
Trust-convergence simulation,  formal RoR analysis & 
No \\
\hline
\cite{Ismail2024Towards} (2024) &
Application &
Device &
ML-based &
Direct & 
T & 
Insider, DoS &
Simulation on WSN-DS dataset &
No \\

\hline
\cite{Li2024Blockchain} (2024) &
Application &
Device &
Bayesian probabilistic trust model based on Beta distribution &
Hybrid & 
E & 
BMA, OOA, SFA, SA &
Simulation &
Available upon request \\

\hline
\cite{Chen2024UITDE} (2024) &
Application &
Device \& Data &
UAV ground-truth accuracy, EWMA decay, Gaussian-based data rectification &
Reported data, per-timestep/historical accuracy &
T &
FDI, DP, collusion &
Experiments on 3 synthetic MCS datasets & 
\href{https://github.com/zchengchen/UITDE}{Yes} \\ 
\hline
\cite{Li2023A} (2023) &
Application &
Device &
Challenge-based &
Hybrid &
E &
Betrayal, SOOA, PMFA &
Simulation, real network &
No \\
\hline
\cite{Yang2023Blockchain} (2023) &
Application &
Device \& Service &
Dirichlet reputation, time decay, punishment-revocation, direct/indirect trust &
Service ratings, rating freshness, interaction success/failure, reputation &
E &
False messages, BMA, OOA &
Simulation on real-world Chongqing taxi GPS dataset  &
No \\
\hline
\cite{Wang2025Trust} (2025) &
Application \& Physical &
Device \& Data &
Weighted multi-dimensional trust with TOPSIS & 

Direct &
T &
Malware/virus propagation, deceptive node attacks, DoS &
Simulation on MATLAB, OMNeT++& 
No \\


\bottomrule
\end{tabular} 
}
\footnotesize{E = Event-driven Update, T = Time-driven Update, E \& T = Hybrid, - = Not Reported}

\end{table*}

Lastly, research in infrastructure IoT trust are given in Table \ref{tab:taxonomy_comparison-infrastructure1} and Table \ref{tab:taxonomy_comparison-infrastructure2}. This domain is the most device-centric domain in the reviewed set, with device trust appearing in 95\%. However, unlike Industrial IoT, infrastructure studies more often combine multiple trust scope. Assessing device and data together 26\%, reflects the need to validate both participating entities and mobility-generated observations in large-scale deployments. The evaluation focus also shifts downward to the network layer 53\% of studies, where trust supports secure routing, relay selection, clustering, virtual-network embedding, and grid communication \cite{Li2024Blockchain, Kaur2024DRIVE, Ahmed2022Link, Mohanta2025Smart, Han2024A}. \cite{Ahmed2022Link} introduces link-history-based trust penalization as a load-balancing mechanism 
which is useful for routing/forwarding trust and edge-security sections. In addition to routing decisions, some frameworks further integrate trust with network resource management. For example, \cite{Ahmed2022Link} incorporates link-history-based trust into load-balancing decisions, demonstrating that trust can also mitigate congestion and denial-of-service conditions by avoiding overloaded forwarding paths.

\begin{table*}[h] 
\centering
\caption{Continue Tab. 6}
\label{tab:taxonomy_comparison-infrastructure2}
\resizebox{\textwidth}{!}{%
	
\begin{tabular}{p{0.8cm} p{1.5cm} p{1.2cm} p{2.9cm} p{3.6cm} p{0.8cm} p{2.6cm} p{3cm} p{1.2cm} }
\toprule

\multicolumn{2}{c}{\textbf{Research Scope}} & 
\multicolumn{4}{c}{\textbf{Trust Design}} & 
\multicolumn{3}{c}{\textbf{Evaluation \& Security}}\\
\cmidrule(r){1-2} \cmidrule(lr){3-6} \cmidrule(lr){7-9} 

\textbf{Ref. (Year)} & 
\textbf{Evaluation Layer} & 
\textbf{Trust Target} &
\textbf{Computation Model} & 
\textbf{Trust Inputs} & 
\textbf{Update} &
\textbf{Attacks Mitigated} & 
\textbf{Validation} & 
\textbf{Code?}  
\\ 
\midrule


\cite{Bounaira2025Trustworthy} (2025) &
Network &
Device &
Game theory & 
Interaction history, reputation, location, resources &
E \& T & 
Malicious nodes interference, selfish behavior &
Simulation, hedonic coalition, Stackelberg games &
Yes \\

\hline
\cite{Mohanta2025Smart} (2025) &
Network &
Device &
Probabilistic  &
Digital signatures, ID credentials, energy usage data, timestamps &
E \& T &
Single point of failure, unauthorized access, data tampering, privacy leaks &
Ethereum testbed &
No \\

\hline
\cite{Kaur2024DRIVE} (2024) &
Network &
Device &
Weighted sum, forgetting factor, reward/penalty &
Packet-forwarding ratio, recommendations &

E &
Malicious nodes (false reports, misbehaving CH) & 
Simulation on Python, Freeway mobility &
No \\

\hline
\cite{Li2024Energy} (2024) &
Network &
Device & 
Interaction success/failure &
Historical interaction success/failure, real-time task-execution outcomes, node attributes &
E \& T &
Malicious edge nodes, network/data/migration disruption, errors &

Simulation on Java, NetLogo & 
No \\
\hline
\cite{Zhang2024Security} (2024) &
Network &
Device &
Weighted sum, DRL for trust-constrained VNE&
Security level, node activity, communication status, VNR participation &
E &
Malicious host node, data tampering/leakage &
Simulation on NetworkX  &
No \\
\hline
\cite{Zhang2024Towards} (2024) &
Network &
Device \& Service &
Subjective logic, discounting/consensus, DQN energy prediction &

Ownership, honesty, social relationships, latency, PDR &

E \& T &
Collusion, GMA, OOA, malicious recommendations &
Simulation on MATLAB, vs TwI-FTM, TTLA, MTTM & 
No \\
\hline
\cite{Ahmed2022Link} (2022) &
Network &
Device &
Beta distribution, weighted direct+cumulative trust, recommendation aggregation &

Communication behavior, energy, data similarity, link history, recommendations &
E \& T &
DoS, energy-based node compromise &
Simulation on MATLAB, vs BLTM & 
No \\

\hline
\cite{Dehalwar2022Blockchain} (2022) &
Network &
Device \& Data &
Blockchain authentication, Merkle-tree/SHA-256 verification &

Identity credentials, signatures, block hashes, consensus votes, timestamps &
E &
Identity theft, masquerading, SA, FDI &
PoC with synthetic data &
No \\
\hline
\cite{Haseeb2022Trust} (2022) &
Network &
Device \& Data &
Weighted direct trust &
Link quality, delay, energy, relay trust, distance, mobility &
T &
FDI, malicious node attacks, confidentiality/auth threats &
Simulation on NS-3, vs fuzzy-IoT, CPSLP & 
No \\
\hline
\cite{Qureshi2022Nature} (2022) &
Network &
Device &
Weighted hybrid trust, differential evolution fitness &

Packet forwarding, availability, residual energy, connectivity, queue congestion &
T &
BHA &
Simulation on NS-2, vs LEACH, TMS, eeTMFO/GA &
No \\

\hline
\cite{Han2024A} (2024) &
Network \& Physical &
Device \& Data &
RL Q-learning self-trust, signature-based backtracking &
Energy, activity, interaction frequency-success rate, data reliability, signatures &
E \& T &
NC, data tampering, SFA &
Simulation &
No \\

\bottomrule
\end{tabular} 
}
\footnotesize{E = Event-driven Update, T = Time-driven Update, E \& T = Hybrid, - = Not Reported}

\end{table*}

Trust inputs are strongly mobility and resource-aware, using packet-forwarding behavior, interaction history, energy, link reliability, and location evidence. Computationally, infrastructure frameworks frequently couple trust with optimization or game-theoretic decision-making \cite{Li2024Energy, Qureshi2022Nature, Bounaira2025Trustworthy, Zhang2024Security}. FDI remains the most common attack by 37\%, but reputation and forwarding attacks are also visible in vehicular and routing frameworks \cite{Xiang2025Secure, Li2024Blockchain, Yang2023Blockchain, Qureshi2022Nature}. Overall, the shared goal of infrastructure IoT trust management is to preserve dependable large-scale operation under mobility, multi-stakeholder governance, and critical-service availability constraints.

In summary, Table \ref{tab:cross-domain-comparison} compares four application domains into a side-by-side comparison of how the surveyed frameworks differ in edge placement, evaluation layer, trust target, trust evidence acquisition, computation method, update pattern, and reproducibility aspects.
For the edge utilization, studies in commercial and industrial IoT most often place trust computation at MEC brokers, service gateways, or Edge-AI nodes. Because service ranking and process-data filtering are latency-sensitive. Studies in consumer IoT have shifted toward federated and device-local training to keep household and health evidence off a central server. Infrastructure frameworks, despite our edge-enabled scope, more often evaluate trust on the forwarding path itself; UAV, RSU, or load-balancing edges appear, but many routing schemes remain node-local.

Trust targets follow the operational object of each domain. Device trust dominates everywhere, yet commercial IoT is the only domain in which service trust is the most common design goal (36\% versus 17\%, 10\%, and 11\%). Data trust is rarely standalone: it is evaluated together with device trust in consumer IoT, used as a process-integrity check in industrial IoT, and paired with mobility observations in 26\% of infrastructure IoT studies. 
In trust evidence acquisition,  consumer and commercial IoT research typically take into consideration direct observations with recommendations, industrial IoT schemes weight interaction feedback at the edge whereas infrastructure IoT schemes rely more on direct network observables such as packet delivery, energy, and link quality.
Adaptive computation, meaning models that revise weights or thresholds from observed behavior rather than from a fixed formula, reaches about half of consumer and commercial IoT studies. This includes federated learning, DRL offloading, fuzzy inference based works. But it is less common in industrial (40\%) and infrastructure IoT (32\%), where weighted, probabilistic, and game-theoretic rules still critical choice. 
Event-driven or hybrid updates are the majority pattern in all four domains; purely periodic updates remain a minority. 
This survey reveals that only three of 55 studies release public code, and seven more offer it upon request. Therefore, cross-paper comparison still depends on incompatible simulators and author-built traces.

\begin{table*}[t]
\centering
\footnotesize
\caption{Cross-Domain Comparison of Characteristic Trust-Design Choices ($N=55$)}
\label{tab:cross-domain-comparison}
\begin{tabular}{p{1.4cm} p{2.8cm} p{3cm} p{3cm} p{3.1cm}}
\toprule
\textbf{Design  Item} & \textbf{Consumer IoT} 

($N{=}12$) & \textbf{Commercial IoT} 

($N{=}14$) & \textbf{Industrial IoT} 

($N{=}10$) & \textbf{Infrastructure IoT} 

($N{=}19$) \\
\midrule
Edge 

Utilization &
Federated or local training, cloud-edge reputation. &
Highest fog/MEC placement, service brokers and offloading. &
Edge filters feedback, Edge-AI and digital twins. &
Path-level computation, UAV/RSU edges, often node-local. \\
\midrule
Evaluation Layer &
App 33\%, cross-layer 33\%, net 25\%. &
Net 50\% and app 43\%, service/device split. &
App 60\%, gates industrial control. &
Net 53\% for routing and clustering, app 37\%. \\
\midrule
Trust 

Target &
Device 92\%, data 42\%, service 17\%. &
Device 71\%, service 36\%, data 29\%. &
Standalone device 70\%, service 10\%. &
Device 95\%, data 26\%. \\
\midrule
Trust 

Evidence &
Direct \& recommendation, link and context signals. &
Hybrid and QoS/SLO attributes and delay. &
Edge-aggregated feedback, data quality as device proof. &
Mobility and resource metrics, PDR, energy, location. \\
\midrule
Adaptive 

Computation &
50\% learning-based, weighted aggregation. &
50\% with DRL, FL, 

MCDA alternatives. &
40\% for anomaly detection and sensor rejection. &
Lowest at 32\%, defaults to Beta/Dirichlet and game theory. \\
\midrule
Trust 

Update &
Hybrid 42\%, event or time 25\% each. &
Event 43\%, time 36\%, hybrid 21\%. &
Event and hybrid 40\% each, transaction-driven. &
Event 42\%, hybrid 32\%, mobility-driven handover. \\
\midrule
Code 

Availability &
None public, 2 upon request, simulation-heavy. &
1 public, 1 upon request, simulator-heavy. &
None public, 3 upon request, some prototypes &
2 public, 1 upon request, simulation and Ethereum tests. \\
\bottomrule
\end{tabular}
\end{table*}

\subsection{Roles of Trust in IoT Operations}\label{sec-roles}

\begin{figure}[h]
    \centering
    \includegraphics[width=0.6\linewidth]{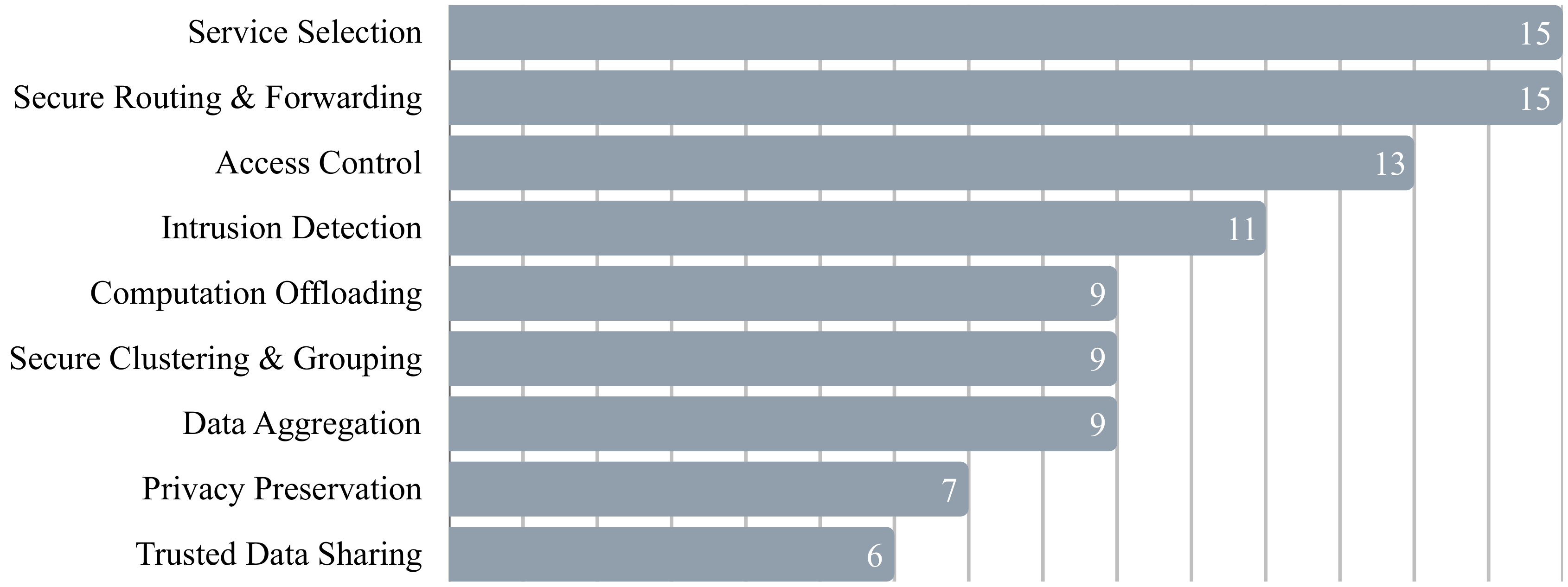}
    \Description{Bar chart showing the distribution of operational decisions supported by trust management in the surveyed literature.}
    \caption{Distribution of Operational Decisions Supported by Trust Management in the Surveyed Literature}
    \label{fig:trust_decision_bar}
\end{figure}

We investigate many state-of-art studies focusing how trust values obtained and computed. It is equally important to examine how these trust scores are actually utilized in operational decision-making. Instead of just saying a device, data or service is trusted, we need to map that trust to specific actions.
This work surveys representative trust-supported IoT operations, summarized in Table~\ref{tab:trust_decisions} and illustrated in Figure \ref{fig:trust_decision_bar}.

\begin{table*}[ht]
\centering
\caption{Taxonomy of Trust-Supported Decisions in IoT Operations }
\label{tab:trust_decisions}
\footnotesize
\setlength{\tabcolsep}{4pt}
\renewcommand{\arraystretch}{1.12}
\begin{tabular}{@{} p{2.5cm} p{3.5cm} p{3.5cm} p{3.5cm} @{}}
\toprule
\textbf{Category} &
\textbf{Trust-Guided Question} &
\textbf{Operational Action} &
\textbf{References} \\
\midrule

Service Selection &
Which provider should be invoked? &
Choose service instance or supplier &

\cite{Wang2025A, Huang2025Towards, Bampatsikos2025Trust, Bangui2025Leveraging, Tsekenis2025Flexible, Yu2024Dynamic, Sun2024Trust, Liang2024Collaborative, Souri2024A, Feng2023Blockchain, Yu2023CET, Wang2023EIDLS,Guo2022Endogenous,Qureshi2022Nature, Lu2025Towards} \\ 
\addlinespace
\hline
Secure Routing \& Forwarding &
Who should relay packets next? &
Select next hop or path &

\cite{Hassan2024LETM, Sharma2025Evaluation, Deng2024Lightweight, farooq2022multi, airehrour2019sectrust, Gheisari2025A, Wang2024Secure, Khan2024Trust, Zhang2024Towards, Kaur2024DRIVE, Ahmed2022Link, Reddy2025AI, Han2024A, Sharma2024Multi, Haseeb2022Trust}  \\ 
\addlinespace
\hline
Access Control &
Should the entity be allowed in? &
Grant, deny, or revoke access &

\cite{Zhang2024Towards, Ahmad2025TrustAware, Ahmed2025Trust, Padmavathi2025A, Xiang2025Secure, Wang2024Secure, Ali2024TIHCS, Egala2023Fortified, Ajao2023Secure,Dehalwar2022Blockchain,Wazid2022TACAS, Latif2022A,Fang2020Fast} \\ 
\addlinespace
\hline
Intrusion Detection &
Is the entity malicious? &
Detect, classify, or isolate misbehavior &

\cite{Anwer2025TEAD, Rehman2025A, Han2024A, Rathee2025An, Ali2024TIHCS,Li2023Federated, Moudoud2022Detection, Verma2023RepuTE, Li2020A, Wang2021Mobile,Alsheakh2020Towards}  \\ 
\hline
Computation Offloading &
Where should the task execute? &
Offload to local, edge, or cloud node &

\cite{Bounaira2025Trustworthy, Zhang2024Security, Li2024Energy, Liang2024Collaborative, Chen2023A,Faraji2024A, Souri2024A, Zhu2024A, Ahmed2022Link}  \\ 
\addlinespace
\hline
Secure Clustering \& Grouping &
Which nodes form a stable team? &
Form clusters; elect cluster head &

\cite{Messina2025Forming, Xiang2025Privacy,Tuncel2025SAFE,Kuppusamy2024IoT,Kaur2024DRIVE, Fortino2023A,Patnaik2023Blockchain, Qureshi2022Nature, Salim2022SEEDGT} \\ 
\addlinespace

\hline
Data Aggregation &
Which reports should be fused? &
Accept, reject, or weight sensor data &

\cite{Wang2025Trust, Khan2025An, Sharma2024Multi, Moudoud2022Detection, Guo2022ITCN, Salim2022SEEDGT, AlMaslamani2022Secure, Zhang2021AIT, Ratnayake2024Machine} \\ 
\addlinespace

\hline
Privacy Preservation &
What must stay hidden? &
Anonymize or restrict sensitive disclosure &

\cite{Li2024Blockchain,Zehra2024A, Osamy2025TAGSCS, Gheisari2025A, Gnanajeyaraman2023VANET,Le2022Artificial, Ratnayake2024Machine} \\ 
\addlinespace

\hline
Trusted Data Sharing &
To whom should data be sent? &
Select trusted relay, peer, or cloud endpoint &

\cite{Khan2025Trust, Wang2025Attribute, Kumar2025AI,Gnanajeyaraman2023VANET,ElMajdoubi2020Towards, Kuppusamy2024IoT}  \\ \addlinespace

\bottomrule
\end{tabular}
\end{table*}

The obtained results show that trust is most frequently integrated into routing and forwarding and as well service selection decisions, indicating that trust management is primarily used to regulate which entities an IoT node should interact with. Access control and intrusion detection are also prominent, extending this role from selecting communication partners to deciding whether an entity should be admitted or considered malicious. Together, these categories show that current frameworks use trust as a preventive or selective mechanism for controlling interactions with potentially unreliable IoT entities. In contrast, trust is less frequently incorporated into computation offloading, secure clustering, and data aggregation. These applications require trust to be combined with additional operational objectives such as resource allocation, energy efficiency, data quality, or group formation, suggesting that trust is less often treated as the primary decision criterion when multiple system-level objectives must be optimized. Data-oriented decisions are particularly limited: trusted data sharing, data aggregation and privacy preservation receive less attention than communication and device-related decisions. This shows that trust is more often used to evaluate devices than the data they provide or share.

\subsection{Attack Mitigation in State-of-the-Art Frameworks}\label{sec-state-attacks}

We examine how state-of-the-art trust management frameworks use trust to mitigate attacks. It is equally important to identify which attacks these frameworks address and how broadly their security mechanisms are evaluated. Table~\ref{tab:attacktable} summarizes the attacks considered by the surveyed frameworks. 
FDI is the most frequently addressed attack, followed by BMA, GMA, SA, and OOA. Within the literature analyzed by \cite{konsta2022trust}, BMA, GMA, and SPA emerge as the most frequently addressed trust-related attacks, while OOA and OSA receive less attention. These attacks affect the evidence used to assess entities or directly alter their perceived behavior. DP is also repeatedly considered particularly in learning-based and data-driven trust mechanisms.

\begin{table}[h] 
\centering
\caption{Mapping of Studies to Mitigated Trust-Related Attacks}
\label{tab:attacktable}
\footnotesize

\begin{tabular}{p{2cm} c c c c c c c c c c c c c c c c}
\toprule
\textbf{Study} 
& \textbf{NC} 
& \textbf{SDA} 
& \textbf{SA} 
& \textbf{FDI} 
& \textbf{JA} 
& \textbf{BMA} 
& \textbf{GMA} 
& \textbf{SPA} 
& \textbf{DP} 
& \textbf{OOA} 
& \textbf{SFA} 
& \textbf{BHA} 
& \textbf{GHA} 
& \textbf{OSA} 
& \textbf{DA} 
& \textbf{WA} 
\\
\midrule
\cite{Ahmad2025Adaptive} &-- &\checkmark &\checkmark &-- &-- &-- &-- &-- &-- &\checkmark &-- &-- &
-- &-- &\checkmark & -- \\
\hline
\cite{Ahmed2025Trust} &-- &-- &\checkmark &-- &-- &-- &-- &-- &-- &-- &-- &-- &-- &-- &\checkmark & -- \\ 
\hline
\cite{Anwer2025TEAD} &-- &-- &\checkmark &\checkmark &-- &-- &-- &-- &-- &-- &\checkmark &-- &-- &
\checkmark &\checkmark & -- \\
\hline
\cite{Chen2023A} & -- & -- & -- & -- & \checkmark & -- & -- & -- & -- & -- & -- & -- & -- & -- & -- & -- \\

\hline
\cite{Chen2024UITDE} & -- & -- & -- & \checkmark & -- & -- & -- & -- & \checkmark & -- & -- & -- & -- & -- & -- & -- \\
\hline
\cite{Han2024A} & \checkmark & -- & -- & \checkmark & -- & -- & -- & -- & -- & -- & \checkmark & -- & -- & -- & -- & -- \\
\hline
\cite{Islam2025Decentralized} &-- &-- &\checkmark &\checkmark &-- &-- &-- &-- &-- &-- &-- &-- &-- &
\checkmark &-- & -- \\
\hline
\cite{Ismail2024Towards}      & -- & -- & \checkmark & -- & -- & -- & -- & -- & -- & -- & \checkmark & -- & -- & -- & -- & -- \\ 
\hline
\cite{Khoshvaght2025An} &-- &-- &-- &\checkmark &-- &\checkmark &\checkmark &\checkmark &-- &-- &-- &-- &-- &-- &\checkmark & -- \\ 
\hline
\cite{Kumar2025AI} & \checkmark & -- & \checkmark & -- & -- & -- & -- & -- & -- & -- & -- & -- & -- & -- & \checkmark & --  \\ 
\hline
\cite{Le2022Artificial} & -- & -- & \checkmark & \checkmark & -- & -- & -- & -- & \checkmark & -- & -- & -- & -- & -- & -- & -- \\
\hline
\cite{Li2023Federated} & -- & -- & \checkmark & -- & -- & -- & -- & -- & -- & \checkmark & -- & -- & -- & -- & -- & -- \\
\hline
\cite{Mahmood2025Application}   & -- & -- & -- & -- & -- & \checkmark & \checkmark & -- & -- & -- & -- & -- & -- & -- & -- & -- \\ 
\hline

\cite{Moeinaddini2025A} & -- &  \checkmark & -- & --  & -- & \checkmark & \checkmark & \checkmark & -- &  \checkmark & -- & -- & -- & \checkmark & \checkmark  & -- \\
\hline

\cite{Patnaik2023Blockchain} & -- & -- & \checkmark & -- & -- & \checkmark & \checkmark & -- & -- & -- & -- & -- & -- & -- & -- & \checkmark \\
\hline

\cite{Ranathunga2021The} & -- & -- & -- & -- & \checkmark & -- & -- & -- & -- & -- & -- & \checkmark & -- & -- & -- & -- \\

\hline

\cite{Rehman2025A} &\checkmark &-- &-- &-- &-- &\checkmark &\checkmark &-- &-- &\checkmark &-- &-- &-- &-- &\checkmark & -- \\
\hline
\cite{Sharma2024Multi} & -- & -- & -- & \checkmark  & -- & \checkmark & -- & -- & -- & -- & -- & -- & \checkmark & -- & -- & -- \\
\hline
\cite{Verma2023RepuTE} & -- & -- & \checkmark & -- & -- & -- & -- & \checkmark & -- & -- & -- & -- & -- & -- & -- & \checkmark \\
\hline
\cite{Wang2021Mobile} & -- & -- & -- & \checkmark & -- & \checkmark & -- & -- & -- & -- & \checkmark & -- & -- & -- & -- & -- \\
\hline

\cite{Xiang2025Secure} & \checkmark & -- & -- & \checkmark & -- & -- & -- & -- & -- & -- & -- & -- & -- & -- & -- & -- \\
\hline

\cite{Zehra2024A} & -- & -- & \checkmark & \checkmark & \checkmark & -- & -- & -- & -- & -- & -- & -- & -- & -- & -- & -- \\
\hline

\cite{Moudoud2022Detection, Bampatsikos2025Trust}   & -- & -- & -- & \checkmark & -- & \checkmark & \checkmark & \checkmark & \checkmark & -- & -- & -- & -- & -- & -- & -- \\ 
\hline
\cite{Qureshi2022Nature, Salim2022SEEDGT} & -- & -- & -- & -- & -- & -- & -- & -- & -- & -- & -- & \checkmark & -- & -- & -- & -- \\
\hline
\cite{Zhang2024Towards, Fortino2023A} & -- & -- & -- & -- & -- & \checkmark & \checkmark & \checkmark & -- & \checkmark & -- & -- & -- & -- & -- & -- \\ 
\hline
\cite{Yang2025PRADA, Dehalwar2022Blockchain} & -- & -- & \checkmark & \checkmark & -- & -- & -- & -- & -- & -- & -- & -- & -- & -- & -- & -- \\ 
\hline
\cite{Yu2024Dynamic,Li2020A} & -- & -- & -- & -- & -- & \checkmark & \checkmark & -- & -- & \checkmark & -- & -- & -- & -- & -- & -- \\ 
\hline
\cite{Yu2023CET, Fang2020Fast} & -- & -- & -- & \checkmark & -- & -- & -- & -- & -- & \checkmark & -- & -- & -- & -- & -- & -- \\
\hline

\cite{Yang2023Blockchain,Zhang2021AIT} & -- & -- & -- & \checkmark & -- & \checkmark & -- & -- & -- & \checkmark & -- & -- & -- & -- & -- & -- \\ 
\hline
\cite{Ajao2023Secure,Singh2023TaLWaR,Sylla2021SETUCOM} & \checkmark & -- & \checkmark & \checkmark & -- & -- & -- & -- & -- & -- & -- & -- & -- & -- & -- & -- \\
\hline
\cite{Khan2025An, AlMaslamani2022Secure, Wang2023EIDLS} & -- & -- & -- & -- & -- & -- & -- & -- & \checkmark & -- & -- & -- & -- & -- & -- & -- \\
\hline

\cite{Guo2022ITCN, Latif2022A,Wang2020MTES, ElMajdoubi2020Towards, Wang2025Attribute,Haseeb2022Trust, Huang2025Towards} & -- & -- & -- & \checkmark & -- & -- & -- & -- & -- & -- & -- & -- & -- & -- & -- & -- \\

\bottomrule
\end{tabular}
\end{table}

The remaining attacks NC, JA, SFA, BHA, GHA, OSA, DA, and WA appear in only a small number of studies. This does not directly indicate that they are less relevant to trust management. But, for example, NC can change the trust status of an entity after compromise. JA can affect the evidence available for trust assessment. SFA, BHA, and GHA can directly alter the observed forwarding behavior of an entity. Similarly, OSA, DA, and WA can manipulate service behavior or the historical evidence associated with an entity. Thus, the reviewed attacks can all be interpreted as threats to the trust assessment, trust evidence, or trust-driven decision process. The main difference is where they affect this process and which trust evidence they manipulate. Therefore, we studied this issue in Table \ref{tab:attack-taxonomy-further}. A notable gap is that relatively few frameworks evaluate several of these attack types together. For example, \cite{Moeinaddini2025A} addresses SDA, BMA, GMA, SPA, OOA, OSA, and DA, while \cite{Moudoud2022Detection, Bampatsikos2025Trust} jointly consider FDI, BMA, GMA, SPA, and DP. However, this shows that recent frameworks increasingly evaluate trust under multiple forms of adversarial behavior rather than a single attack model.

\section{The Road Ahead: Building a Trustworthy IoT Ecosystem} \label{sec-road-ahead}

As IoT devices increasingly make autonomous, localized decisions at the edge, trust management can no longer rely on simple scalar values. It must consistently evaluate devices, data, and services across the edge-cloud continuum while withstanding mobility, resource constraints, and trust-related attacks. Although emerging paradigms like edge AI, federated learning, blockchain, and 6G enable distributed trust, they also introduce novel complexities. 

\subsection{Research Challenges} \label{sec-road-challenges}
\begin{figure}[h]
    \centering
    \includegraphics[width=0.98\linewidth]{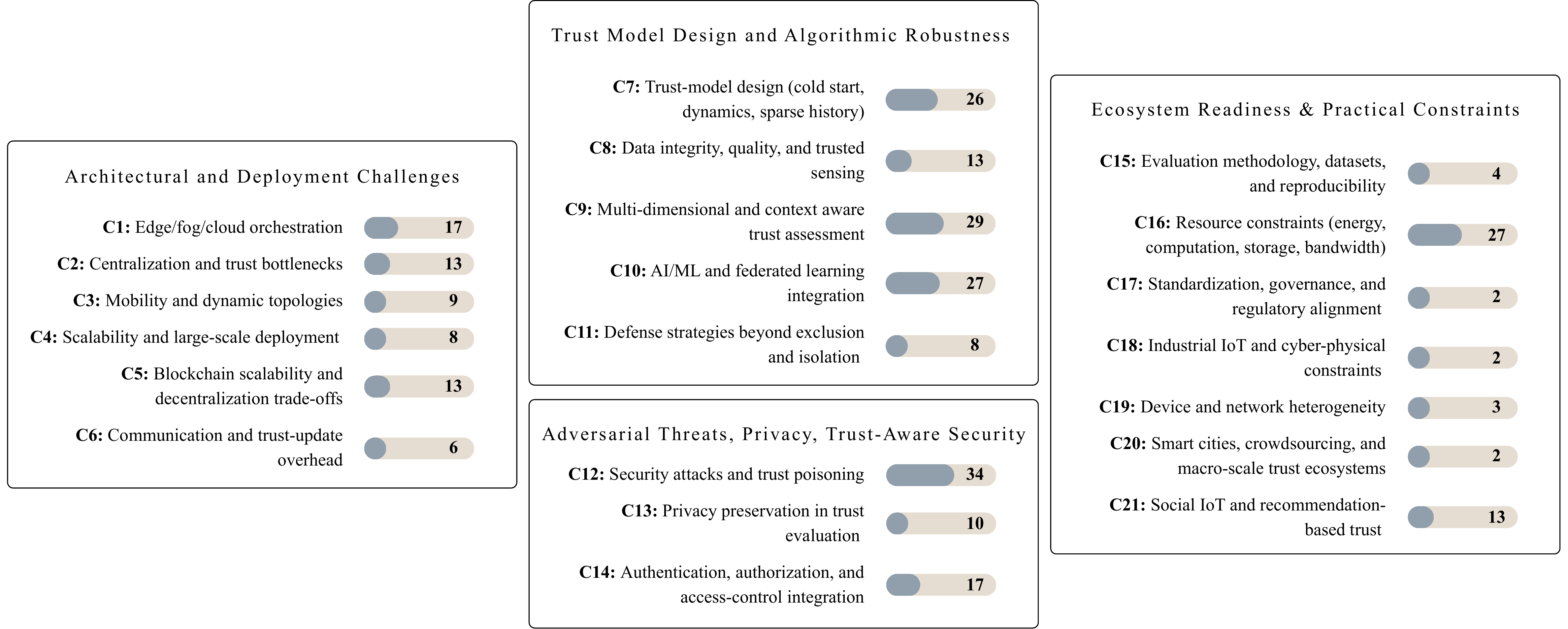}
    \Description{Frequency distribution of research challenges in surveyed literature.}
    \caption{Frequency Distribution of Specific Research Challenges Identified in the Surveyed Literature (N=55)}
    \label{fig:challenge-groups}
\end{figure}

In order to critically evaluate the current state of the art and map the trajectory of trust management research, we review the recurring limitations in trust-model design, security, deployment, and ecosystem readiness, as summarized in Figure~\ref{fig:challenge-groups}.

\subsubsection{Architectural and Deployment Challenges}
\label{sec-challenge-architecture}
Maintaining consistent trust across distributed edge, fog, and cloud infrastructures (C1) has gained increasing attention over the years. It emerges as the most cited architectural challenge in the state-of-art studies. This challenge has grown from a minority concern in 2020–2021 
to a central focus by 2025 
as edge AI, digital twins, and cloud--edge--terminal architectures become the dominant deployment paradigm~\cite{Khan2025An, d2025blockchain, konsta2022trust, Ranathunga2021The}. The challenge here is not just the physical placement of computing, but also ensuring the consistency of trust state across these different tiers. 
Closely related to this challenge, the degree of centralization that trust management should adopt is another important research challenge (C2). This challenge arises because fully distributed peer-based trust mechanisms converge slowly and rely on sparse local observations, whereas centralized authorities provide globally consistent trust decisions at the cost of higher latency and single points of failure~\cite{Gnanajeyaraman2023VANET}.
Beyond maintaining trust consistency in an appropriate trust management architecture, trust mechanisms must also adapt to constantly changing network topologies. Supporting mobility and dynamic topologies (C3) becomes essential especially in between vehicles, or UAV. Because in mobile networks, trust evidence can quickly become outdated due to handovers, intermittent connectivity, and network partitions. Therefore, maintaining accurate and timely trust assessment is challenging ~\cite{Zhang2024A,farooq2022multi, Reddy2025AI}. Trust management in IoT must also remain efficient at large scale. Ensuring scalability (C4) requires controlling the communication, computation, and storage overhead introduced by trust management as IoT deployments grow. The first thing that comes to mind when scalability is addressed is generally the size of the network. However, it is important to emphasize that scalability issues are also due to various types of traffic, topology changes due to mobility, and the need to perform trust evaluation, routing, and quality of service simultaneously \cite{Reddy2025AI, Xiong2023BDIM}. In the research by \cite{konsta2022trust}, it is noted that IoT nodes may accumulate large volumes of trust-related information, yet storage complexity analysis is almost absent.

One frequently proposed response to the scalability limitations of trust management systems is the use of blockchain-based infrastructures. Evaluating blockchain trade-offs (C5) therefore becomes an important challenge on its own. While blockchain can provide tamper-evident trust records and decentralized consensus mechanisms~\cite{Egala2023Fortified, jmal2025blockchain,konsta2022trust,Feng2023Blockchain, Ismail2024Towards}, its practical deployment introduces additional overhead through ledger growth, consensus latency, and on-chain trust-state storage. These costs are often difficult to accommodate on constrained IoT devices, and existing hybrid solutions rarely provide comprehensive cost models spanning computation, communication, and storage resources.

Whenever trust information is shared beyond a single device, the network must carry additional control traffic, including reputation broadcasts, certificate or token exchanges, blockchain transactions, and federated or edge-AI model updates. This overhead consumes the same constrained LPWAN and low-power mesh links used for application data, increasing latency, packet loss, and energy consumption even when the underlying trust algorithm is computationally lightweight ~\cite{Wang2023EIDLS,Wazid2022TACAS,muzammal2020comprehensive}. Although only six studies among surveyed research explicitly identify communication and trust-update overhead (C6) as an open research challenge. Because in state-of-art research many suggestions assume that trust is spread on a regular basis, but they do not report how often messages are sent, how big updates are, or how this affects network overall. 
Maintaining trust is not cost-free. Update propagation add communication and energy overhead to IoT nodes \cite{Yang2023Blockchain}. Nevertheless, current literature often prioritizes the initial trust computation over the development of sustainable, ongoing maintenance mechanisms. 

\subsubsection{Trust Model Design and Algorithmic Robustness}
\label{sec-challenge-trust-model}
Beyond architectural considerations, trust management fundamentally depends on the ability to derive reliable trust scores from incomplete and changing evidence. Designing robust trust models (C7) therefore remains a major research challenge. Trust models must cope with cold-start conditions, sparse interaction histories \cite{Guo2022ITCN}, dynamic trust updates, and indirect recommendations while adapting to rapidly changing IoT topologies, workloads, and device roles~\cite{Messina2025Forming, Wang2023TMETA}. 

Trust models often ignore evaluating transmitted information. The difference between device misbehavior and data untrustworthiness is frequently ignored by many research \cite{Kumar2025AI}. ~\cite{Tadj2023On,Xiang2025Secure, konsta2022trust} highlight the data integrity problem (C8). For instance, an IoT device can execute routing protocols perfectly but still send noisy crowdsourced labels, or falsify sensor readings. These issues require data-driven trust assessment. Most schemes still incorrectly infer data quality based solely on a device's network-layer behavior, making systems vulnerable to false data from trusted nodes.
Moreover, trust management should also account for the context in which trust is established. Supporting multi-dimensional and context-aware trust (C9) recognizes that a single trust score is often insufficient for IoT systems. A device may be trustworthy for one task but not for another. Likewise, changes in location, or operating conditions should not always be interpreted as malicious behavior. Although fuzzy and policy-based trust models consider these factors, there is still no common framework for representing and managing context-aware trust across IoT environments~\cite{Rehman2025A, Sylla2021SETUCOM}. Integrating AI/ML and FL (C10) introduces a new set of challenges. AI-based trust models can better detect malicious behavior and adapt to changing network conditions using techniques such as ensemble learning, graph neural networks, and deep reinforcement learning~\cite{Chen2023A, AlMaslamani2022Secure, Le2022Artificial}. However, these approaches are also vulnerable to poisoning attacks, adversarial inputs, and limited model interpretability. Defining trust-aware response policies (C11) extends beyond simply identifying malicious participants. Most existing trust schemes react by excluding or isolating nodes once their trust falls below a threshold. However, this approach is often unsuitable for sparse, mobile, or safety-critical IoT networks, where removing a participant may reduce network connectivity or service availability. More flexible response strategies, such as graduated sanctions, temporary quarantine with recovery, and distinguishing unreliable from intentionally malicious behavior, remain largely unexplored~\cite{Wang2022Towards, Faraji2024A, Ali2024TIHCS}.

\subsubsection{Adversarial Threats, Privacy, and Trust-Aware Security}
\label{sec-challenge-adversarial}

Trust management should also protect the trust process from attacks. Defending against trust-targeted attacks (C12) requires securing different parts of the trust ecosystem. Attackers may target the trust pipeline by manipulating how trust evidence is collected, shared, updated, or aggregated, leading to incorrect trust decisions even when individual devices are not compromised. Although many studies discuss these attacks, most evaluate only a single attack type \cite{Moeinaddini2025A,Reddy2025AI}.

While C12 focuses on protecting the trust process from attacks, trust management should also protect the sensitive information used to establish trust. Preserving privacy (C13) is challenging because trust evaluation relies on sensitive information and may subject to privacy regulations~\cite{Osamy2025TAGSCS, Gheisari2025A}. Techniques such as FL and differential privacy reduce data exposure, but they also limit the information available for trust assessment and remain vulnerable to poisoning attacks \cite{Feng2023Blockchain}. As a result, balancing privacy and trust accuracy remains an open research problem.

The challenge in authentication and access-control integration (C14) in trust management frameworks defines a third barrier at the security boundary: cryptographic authentication does not detect compromised insiders, while behavioral trust alone lacks strong identity binding~\cite{Wazid2022TACAS,Mahmood2025Application,muzammal2020comprehensive,hossain2024holistic}.

\subsubsection{Ecosystem Readiness and Practical Constraints}
\label{sec-challenge-ecosystem}
Although resource constraints (C16), standardization and governance (C17), industrial deployment (C18), heterogeneity (C19), large-scale smart city ecosystems (C20), and Social IoT environments (C21) have long been recognized as fundamental IoT challenges, they continue to be highlighted in recent trust management research. These are not new challenges to the IoT community; however, recent studies continue to identify them as major barriers to deploying practical trust management systems.

In addition to these ecosystem challenges, trust management still lacks reliable evaluation methods. Establishing reproducible evaluation methodologies (C15) remains a persistent challenge, although only a few studies identify it explicitly. Previous surveys report inconsistent benchmarks, incompatible simulators, limited public code, and evaluation settings that differ in topology, attack models, and performance metrics, making fair comparison difficult~\cite{de2023survey,kamble2024research,chandrasekaran2024trust,akli2023survey}. Public trust datasets are also limited, often lacking interaction history or recommendation data.
As a result, many studies prefer relying on synthetic data or custom datasets ~\cite{Wang2021Mobile,Chen2024UITDE,Tadj2023On}.

\begin{figure*}
    \centering
    \begin{subfigure}{0.5\linewidth}
        \centering
        \includegraphics[width=\linewidth]{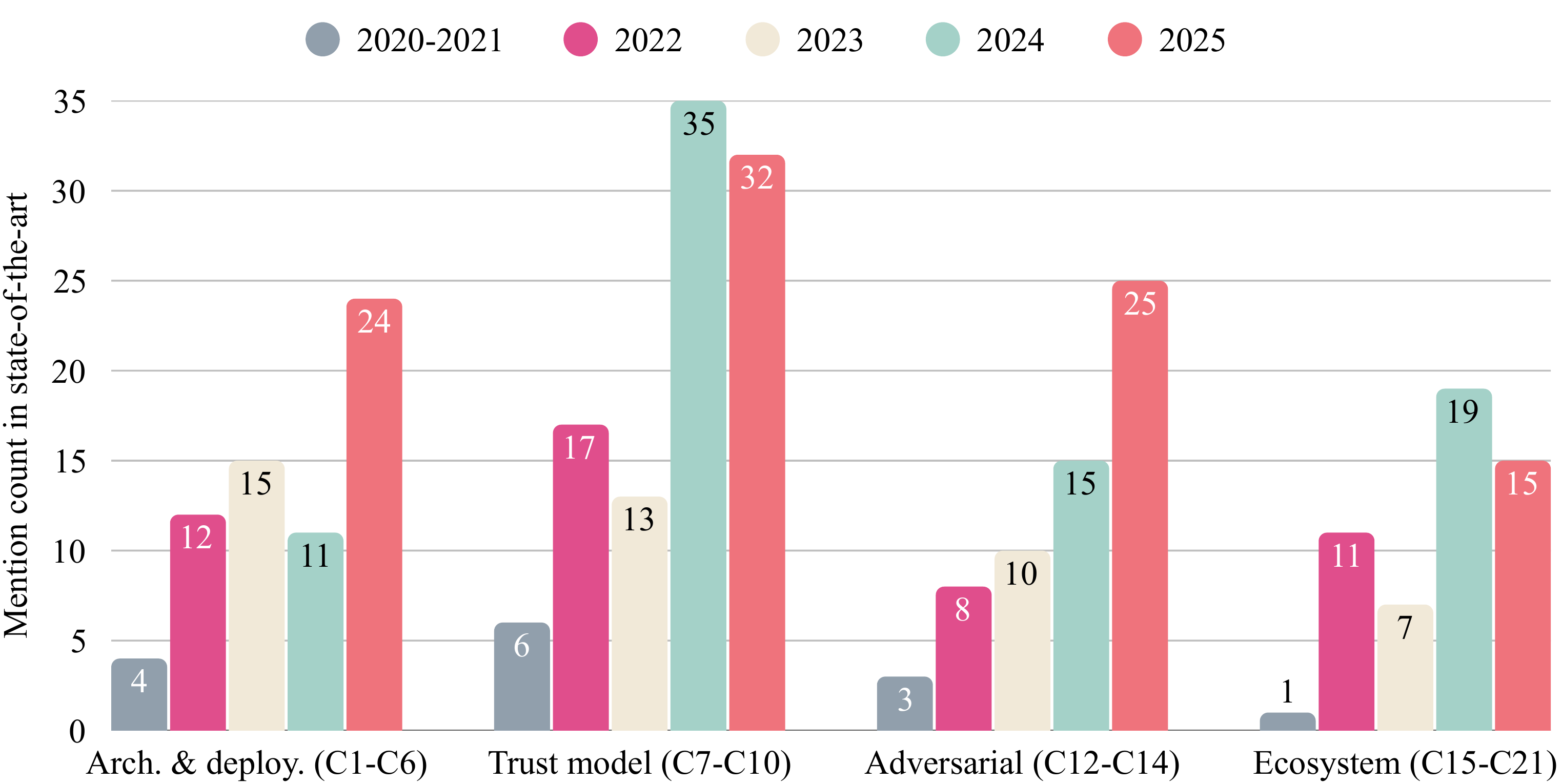}
        \Description{Research Challenges Trends by Years}
        \caption{}
        \label{fig:chart-challenge}
    \end{subfigure}
   \hspace{0.2cm} 
    \begin{subfigure}{0.45\linewidth}
        \centering
        \includegraphics[width=\linewidth]{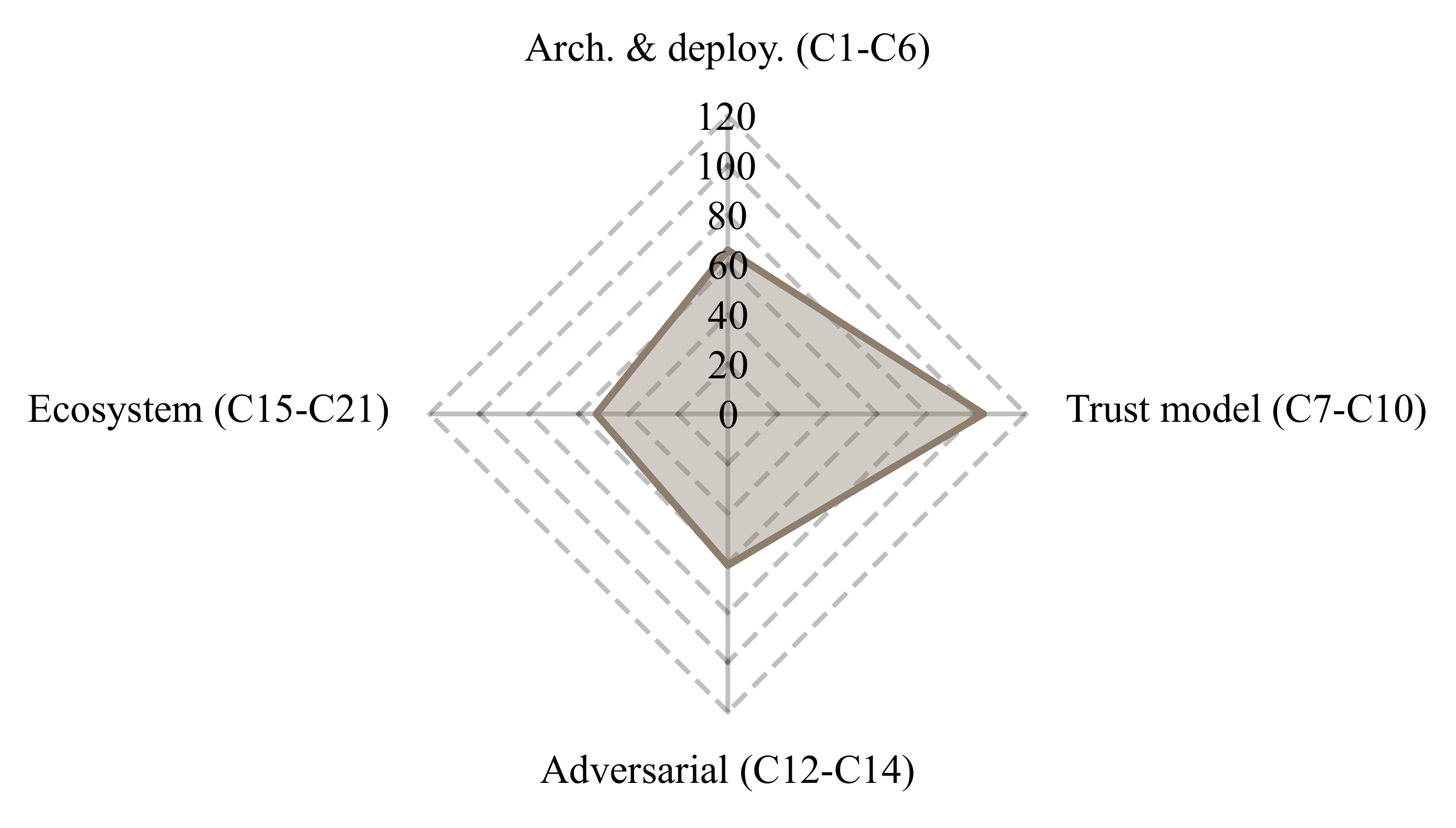}
        \Description{Research Challenges Trends by Broad Areas}
        \caption{}
        \label{fig:spider-challenge}
    \end{subfigure}
    \caption{Research Challenge Trends in the Reviewed Literature by (a) Yearly Distribution and (b) Broad Areas}
    \label{fig:main-challenges}
\end{figure*}

Figure~\ref{fig:main-challenges} groups the 55 surveyed papers into four broad area over 2020-2025. The trend suggests a shift from designing trust models toward deploying and securing them in realistic IoT environments. Trust-model challenges remain dominant, while architectural and adversarial challenges increase notably by 2025. This reflects growing attention to edge AI, distributed trust computation, model poisoning, and other attacks that can manipulate trust evidence or learning processes. In contrast, ecosystem-related challenges show limited growth. Resource constraints are frequently considered, but interoperability, standardization, evaluation, and cross-domain deployment remain less explored. These results indicate that the field is moving beyond trust-score design, while practical questions about how trust mechanisms operate across heterogeneous IoT environments and remain reliable under evolving attacks are still insufficiently addressed.

\subsection{Future Directions} \label{sec-road-future}

After the challenges discussed in Section \ref{sec-road-challenges} what remains open is how the community should tackle them in specific IoT trust designs. Through this perspective, future directions are given below to provide broad technology trends.

\subsubsection{Privacy-Preserving Trust Computation}
Trust calculation requires behavioral and contextual information that may also be valuable to attackers. A central direction is privacy-preserving trust computation, where nodes share abstract trust information instead of raw evidence \cite{Takemura2024TEE}. Semantic trust can reduce direct data exposure by expressing which device capability, data category, or service role is trusted under a given policy; however, semantic representations may still allow sensitive information to be inferred. FL and cryptographic mechanisms could complement semantic trust to further protect the underlying evidence while keeping identifiers, location traces, and rating histories local. 
These kind of mechanisms are specifically valuable in applications in consumer IoT domain.

\subsubsection{Sustainable Trust Management in 6G}
6G introduces new requirements for both sustainability and ubiquitous connectivity. Future trust mechanisms should reduce communication, computation, and energy costs, especially in infrastructure IoT. 
Non Terrestrial Networks, including satellites, UAVs, HAPs, and LEOs can support trust dissemination and computation, but designs must account for mobility, intermittent connectivity, propagation delay, and overhead \cite{Mahmood2025Securing}. Validation must report message rate, update size, latency, and energy per trust score computation and dissemination.

\subsubsection{Agent-to-Agent Trust}
IoT systems are moving from device-centric networks toward autonomous agents, including AI-integrated end devices. A future direction is agent-to-agent trust \cite{Sheng2024Graph}, where agents evaluate one another’s behavior, authority, and reliability from operational evidence rather than identity alone. An emerging extension is to use agents themselves as trust parties: regional trust checkers and managers at the edge could issue short-lived claims, admit or revoke nearby devices, and continuously re-verify peers as their behavior or goals change. Commercial IoT is a favorable choice for this approach, where trust can directly support decisions such as which agent may provide a service, accept a task, or access a resource. However, introducing trust managers also creates a new type of vulnerability: a compromised checker could manipulate trust claims and influence decisions within its region. Future work should therefore examine how such trust authorities can be secured and contained without disrupting the wider network.

\subsubsection{Trust in Virtual, Immersive and Generative IoT Environments} 
Future research should investigate how trust management changes when IoT systems increasingly rely on virtual models and AI-generated information. Digital twins can provide additional evidence for trust assessment by simulating device behavior, testing attack scenarios, and supporting trust bootstrapping for entities even with limited interaction history \cite{Sasikumar2023Blockchain}. However, the trustworthiness of twin-generated evidence itself remains an open question. Similarly, generative AI can support trust prediction and evidence analysis while also introducing new risks through synthetic or manipulated sensor data, identities, and recommendations. Therefore, future trust mechanisms should distinguish between physical and generated evidence and assess their consistency and reliability before using them in trust decisions \cite{Hasan2024NFTs}. This is particularly relevant in large-scale Industrial and Infrastructure IoT, where virtual observations may influence routing, service selection, data aggregation, and other operational decisions. Hybrid physical-virtual testbeds could provide a suitable basis for evaluating these mechanisms under false-data and synthetic-evidence attacks.

\section{Conclusion}\label{sec-conc}

Trust management in edge-enabled IoT requires defining precisely \emph{who} is trusted, for \emph{what} operational decision, and against \emph{which} specific threat. Because prior literature reviews have predominantly restricted their scope to device trust or isolated application domains, this survey adopts a multifaceted analysis to establish a unified perspective. We introduced a comprehensive taxonomy encompassing device, data, and service trust, mapped across a complete trust lifecycle from evidence acquisition to maintenance. Following PRISMA, we systematically analyzed 55 recent studies to evaluate their structural design dimensions across four key IoT domains (consumer, commercial, industrial, and infrastructure) and three architectural layers (application, network, physical).

Our findings indicate that while device-centric frameworks continue to dominate, domain-specific priorities are evolving. Commercial IoT increasingly emphasizes service-level trust, whereas infrastructure deployments frequently integrate device and data trust to support high-mobility environments. In terms of mechanics, trust evidence is primarily derived from interaction histories, reputation metrics, and network observables. Computation models remain largely weighted or probabilistic, though they are increasingly augmented by machine learning and blockchain architectures, with updates handled predominantly through event-driven mechanisms. Consequently, while the research community has largely converged on \emph{how} to compute trust values, a fundamental consensus regarding \emph{what} specific target those scores should represent remains missing. Furthermore, the operational utilization of trust remains significantly narrower than its theoretical design space. Computed trust scores currently function primarily as gatekeepers for basic service selection, secure routing, and access control. Crucial system-level operations such as data aggregation, computation offloading, and privacy-preserving evaluations are rarely integrated, meaning many generated trust metrics are never fully leveraged to drive the decisions that justify their collection. The results for attack coverage reveal that False Data Injection (FDI) emerges as the most frequently addressed vulnerability, followed closely by Sybil attacks (SA), bad-mouthing (BMA), on-off (OOA), and good-mouthing (GMA) attacks. This hierarchy reflects the prevalence of device-oriented schemes, where anomalous data readings are typically penalized as node misbehavior rather than treated as distinct data-trust failures. Meanwhile, routing and service-layer exploits such as black-hole and selective forwarding attacks remain critically underexplored. To design edge-enabled IoT trust management systems into robust, deployable ecosystems, future research must prioritize: (i) maintaining clear, structural distinctions between device, data, and service trust under active attack; (ii) benchmarking trust overheads against measured energy, latency, computation and communication costs; (iii) engineering targeted defenses against decision-flow and resource-exhaustion exploits rather than focusing solely on evidence poisoning; (iv) adopting reproducible evaluation frameworks supported by public datasets and open source code; and (v) validating proposed designs within agent-based, 6G-supported and physically-virtually coupled environments.

\section*{Acknowledgements}
This research was supported by Odine and OdineLabs as part of ImAgSLab Project under the 1773 ITU TTO Project No: 202500303, and by the ZHAW EELISA Sonderfinanzierung SePRIO project.

 \bibliographystyle{ACM-Reference-Format}
 \bibliography{sample-base-titlecase}

\end{document}